\documentclass[12pt]{article}
\usepackage{amsmath,amssymb}
\usepackage{bm}
\usepackage{color}
\usepackage[dvipdfmx]{graphicx}
\usepackage{cite}
\usepackage{mathtools}

\usepackage{multirow}
\usepackage{dcolumn}
\usepackage{longtable}
\usepackage{here}

\def\slash#1{\not\!\!#1}

\begin{document}

\title{
\begin{flushright}
\ \\*[-80pt]
\begin{minipage}{0.2\linewidth}
\normalsize
EPHOU-23-018\\*[50pt]
\end{minipage}
\end{flushright}
{\Large \bf
$Sp(6,Z)$ modular symmetry in flavor structures: \\ 
quark flavor models and Siegel modular forms for $\widetilde{\Delta}(96)$
\\*[20pt]}}

\author{
Shota Kikuchi$^1$
~Tatsuo Kobayashi$^1$,
~Kaito Nasu$^1$,
\\~Shohei Takada$^1$, and
~Hikaru Uchida$^2$
\\*[20pt]
\centerline{
\begin{minipage}{\linewidth}
\begin{center}
{\it \normalsize
$^1$Department of Physics, Hokkaido University, Sapporo 060-0810, Japan \\
$^2$Institute for the Advancement of Graduate Education, Hokkaido University, Sapporo 060-0817, Japan} \\*[5pt]
\end{center}
\end{minipage}}
\\*[50pt]}

\date{
\centerline{\small \bf Abstract}
\begin{minipage}{0.9\linewidth}
\medskip
\medskip
\small
We study an approach to construct Siegel modular forms from $Sp(6,Z)$.
Zero-mode wave functions on $T^6$ with magnetic flux background behave Siegel modular forms at the origin.
Then $T$-symmetries partially break depending on the form of background magnetic flux.
We study the background such that three $T$-symmetries $T_I$, $T_{II}$ and $T_{III}$ as well as the $S$-symmetry remain.
Consequently, we obtain Siegel modular forms with three moduli parameters $(\omega_1,\omega_2,\omega_3)$, which are 
multiplets of finite modular groups.
We show several examples.
As one of examples, we study Siegel modular forms for $\widetilde{\Delta}(96)$ in detail.
Then, as a phenomenological applicantion, we study quark flavor models using Siegel modular forms for $\widetilde{\Delta}(96)$.
Around the cusp, $\omega_1=i\infty$, the Siegel modular forms have hierarchical values depending on their $T_I$-charges.
We show the deviation of $\omega_1$ from the cusp can generate large quark mass hierarchies without fine-tuning.
Furthermore CP violation is induced by deviation of $\omega_2$ from imaginary axis.
\end{minipage}
}
\begin{titlepage}
\maketitle
\thispagestyle{empty}
\end{titlepage}

\newpage


\section{Introduction}
\label{sec:intro}

The origin of both quark and lepton flavor structures are one of challenging issues in the current particle physics.
Experimental results show quarks and charged leptons have large mass hierarchies while neutrinos have extremely small masses.
In addition quarks have small mixing angles while leptons have large one.
Also the CP symmetry is violated in both quark and lepton mass matrices.
These observations are realized in the Standard Model by tuning free parameters and some of them are hierarchical.
Therefore underlying theory explaining the flavor structure naturally is required.

The modular flavor symmetry is one  of attractive ways to approach to the flavor structure.
In modular flavor symmetric models, the Lagrangian has $Sp(2,\mathbb{Z})\simeq SL(2,\mathbb{Z})$ modular symmetry.
Then three-generations of quarks and leptons behave as three-dimensional (reducible or irreducible) representations of the finite modular groups.
Also Higgs fields are representations of  finite modular groups.
Yukawa couplings as well as fermion mass matrices are given by modular forms for  finite modular groups, which are holomorphic functions of the modulus $\tau$ \cite{Feruglio:2017spp} \footnote{The modular flavor symmetry was also studied from the top-down approach such as string theory \cite{Ferrara:1989bc,Ferrara:1989qb,Lerche:1989cs,Lauer:1989ax,Lauer:1990tm,Kobayashi:2018rad,Kobayashi:2018bff,Ohki:2020bpo,Kikuchi:2020frp,Kikuchi:2020nxn,
Kikuchi:2021ogn,Almumin:2021fbk,Baur:2019iai,Nilles:2020kgo,Baur:2020jwc,Nilles:2020gvu,Kikuchi:2023awe}.}.
It is interesting that the finite modular groups $\Gamma_N$ for level $N=2,3,4$ and $5$ are isomorphic to the non-Abelian discrete groups $S_3$, $A_4$, $S_4$ and $A_5$, respectively \cite{deAdelhartToorop:2011re}.
These non-Abelian groups have been used in the flavor models for quarks and leptons \cite{Altarelli:2010gt,Ishimori:2010au,Ishimori:2012zz,Kobayashi:2022moq,Hernandez:2012ra,King:2013eh,King:2014nza,Tanimoto:2015nfa,King:2017guk,Petcov:2017ggy,Feruglio:2019ktm}.
Inspired by this point, the modular symmetric lepton flavor models with $\Gamma_2\simeq S_3$ \cite{Kobayashi:2018vbk}, $\Gamma_3\simeq A_4$ \cite{Feruglio:2017spp}, $\Gamma_4\simeq S_4$ \cite{Penedo:2018nmg} and $\Gamma_5\simeq A_5$ \cite{Novichkov:2018nkm,Ding:2019xna} symmetries have been proposed.
The modular symmetries of higher levels for covering groups were also studied \cite{Li:2021buv,Ding:2020msi,Kobayashi:2018bff,Liu:2019khw,Novichkov:2020eep,Liu:2020akv,Liu:2020msy}.

Phenomenological studies using modular forms for $Sp(2,\mathbb{Z})$ have been implemented in many works \cite{
Criado:2018thu,
Kobayashi:2018scp,
Ding:2019zxk,
Novichkov:2018ovf,
Kobayashi:2019mna,Wang:2019ovr,
Chen:2020udk,
deMedeirosVarzielas:2019cyj,
  	Asaka:2019vev,
Asaka:2020tmo,deAnda:2018ecu,Kobayashi:2019rzp,Novichkov:2018yse,Kobayashi:2018wkl,Okada:2018yrn,Okada:2019uoy,Nomura:2019jxj, Okada:2019xqk,
  	Nomura:2019yft,Nomura:2019lnr,Criado:2019tzk,
  	King:2019vhv,Gui-JunDing:2019wap,deMedeirosVarzielas:2020kji,Zhang:2019ngf,Nomura:2019xsb,Kobayashi:2019gtp,Lu:2019vgm,Wang:2019xbo,King:2020qaj,Abbas:2020qzc,Okada:2020oxh,Okada:2020dmb,Ding:2020yen,Okada:2020rjb,Okada:2020ukr,Nagao:2020azf,Wang:2020lxk,
  	Okada:2020brs,Yao:2020qyy}.
(See fore more references Ref.~\cite{Kobayashi:2023zzc}.)
Some works show deviations of the modulus from the modular fixed points can generate large mass hierarchies without fine-tuning for quarks \cite{Petcov:2022fjf,Kikuchi:2023cap,Abe:2023ilq,Kikuchi:2023jap,Petcov:2023vws,deMedeirosVarzielas:2023crv
} and leptons \cite{Feruglio:2021dte,Novichkov:2021evw}, and for both \cite{Abe:2023qmr,Abe:2023dvr}.
Then the modular forms take hierarchical values depending on their residual charges.
That is, mass matrices written in terms of the modular forms are also hierarchical.
In this sense, the modulus and residual symmetry can be regarded as the origin of the mass hierarchies.

As an another approach to the flavor structures, higher dimensional theories have been studied.
In the theories, the modular symmetry can originate from a geometrical symmetry of the extra dimensions.
For example, the torus compactification $T^2\times T^2\times T^2$ of six extra dimensions in the superstring theory has $Sp(2,\mathbb{Z})\times Sp(2,\mathbb{Z})\times Sp(2,\mathbb{Z})$ modular symmetry as a geometrical symmetry.
Actually some modular forms are derived from the torus compactification of the low-energy effective theory of the superstring theory with magnetic flux background \cite{Kobayashi:2018rad,Kobayashi:2018bff,Ohki:2020bpo,Kikuchi:2020frp,Kikuchi:2020nxn,Kikuchi:2021ogn, Cremades:2004wa}.
Therefore multi modular symmetry such as $\Gamma_{N_1}\times \Gamma_{N_2}\times \Gamma_{N_3}$ can originate from $T^2\times T^2\times T^2$.
Indeed, quark flavors in $\Gamma_6\simeq A_4\times S_3$ \cite{Kikuchi:2023cap} and $S'_4\times S_3$ \cite{Abe:2023ilq} may be derived from the torus compactification with the moduli stabilization $\tau_1=\tau_2\equiv \tau$.
Also multi modular symmetries were studied from phenomenological viewpoint in Refs.~\cite{King:2021fhl,Du:2022lij,Abbas:2022slb}.

Indeed, higher dimensional compact spaces have many moduli and have larger symplectic modular symmetry, $Sp(2g,\mathbb{Z})$, 
which are not direct products of $Sp(2,\mathbb{Z})$.
One of typical string compactificatins is Calabi-Yau compactifications, and they have such larger modular symmetry \cite{Strominger:1990pd,Candelas:1990pi,Ishiguro:2020nuf,Ishiguro:2021ccl}.\footnote{See for symplectic modular symmetry in heterotic orbifold models 
Refs.~\cite{Baur:2020yjl,Nilles:2021glx}.}
Generic tori $T^4$ and $T^6$, which are not direction products of $T^2$, also have larger modular symmetries, 
$Sp(4,\mathbb{Z})$ and $Sp(6,\mathbb{Z})$.
They can lead to a rich flavor structure.
In particular, magnetized toroidal compactification models on higher dimensional torus such as $T^4$ and $T^6$ have been studied \cite{Cremades:2004wa,Antoniadis:2009bg,Kikuchi:2022lfv,Kikuchi:2022psj,Kikuchi:2023awm}.
Magnetized $T^4$ and $T^6$ models have several moduli and have the modular symmetries $Sp(4,\mathbb{Z})\ni Sp(2,\mathbb{Z})\times Sp(2,\mathbb{Z})$ and $Sp(6,\mathbb{Z})\ni Sp(2,\mathbb{Z})\times Sp(2,\mathbb{Z})\times Sp(2,\mathbb{Z})$ as geometrical symmetries.
Then Siegel modular forms for the subgroups of $Sp(4,\mathbb{Z})$ and $Sp(6,\mathbb{Z})$ can be derived from them.

In this paper, we study an approach to construct examples of  Siegel modular forms for  subgroups of $Sp(6,\mathbb{Z})$ from zero-mode wave functions on the magnetized $T^6$.
They are functions of  several moduli; therefore we can expect rich possibilities realizing flavor observations.
Also, as an illustrating example we carry out a numerical analysis of a quark flavor model using the Siegel modular forms, which are constructed by our approach.
We show roles of moduli on the realization of the flavor structure.

This paper is organized as follows.
In section \ref{sec:zero-modes}, we review zero-mode wave functions on the magnetized $T^6$ model.
In section \ref{sec:Sp6Z}, we study $Sp(6,\mathbb{Z})$ modular symmetry.
In section \ref{sec:Examples_Siegel}, we show some examples of the Siegel modular forms of weight 1/2 obtained from the zero-modes on the magnetized $T^6$.
In section \ref{sec:mf}, we study the Siegel modular forms for $\widetilde{\Delta}(96)$ in detail.
In section \ref{sec:Num}, we perform numerical analysis of a $\widetilde{\Delta}(96)$ quark flavor mode with numerical example.
In section \ref{sec:conclusion} is our conclusion.
In appendix \ref{app:group_theory}, we show group theory of $\widetilde{\Delta}(96)$.
In appendix \ref{app:modularforms}, we show the Siegel modular forms for $\widetilde{\Delta}(96)$ up to weight 5.
In appendix \ref{app:ferm}, we summarize fermion mass matrices with $\widetilde{\Delta}(96)$ modular symmetry.


\section{Zero-modes on the magnetized $T^6$ model}
\label{sec:zero-modes}

In this section, we review zero-modes on $T^6$ with background magnetic fluxes \cite{Cremades:2004wa,Antoniadis:2009bg,Kikuchi:2023awm}.
Six-dimensional torus $T^6$ is defined by $T^6\simeq \mathbb{C}^3/\Lambda$, and $\Lambda$ is a six-dimensional lattice spanned by six basis vectors,
\begin{align}
&\vec{e}_1 =
\begin{pmatrix}
1 \\ 0 \\ 0 \\
\end{pmatrix}, \quad
\vec{e}_4 = \Omega \vec{e}_1 =
\begin{pmatrix}
\omega_1 \\
\omega_4 \\
\omega_6 \\
\end{pmatrix}, \\
&\vec{e}_2 =
\begin{pmatrix}
0 \\ 1 \\ 0 \\
\end{pmatrix}, \quad
\vec{e}_5 = \Omega \vec{e}_2 =
\begin{pmatrix}
\omega_4 \\
\omega_2 \\
\omega_5 \\
\end{pmatrix}, \\
&\vec{e}_3 =
\begin{pmatrix}
0 \\ 0 \\ 1 \\
\end{pmatrix}, \quad
\vec{e}_6 = \Omega \vec{e}_3 =
\begin{pmatrix}
\omega_6 \\
\omega_5 \\
\omega_3 \\
\end{pmatrix}.
\end{align}
Here we have defined the complex structure moduli on $T^6$, $\Omega$, as
\begin{align}
\Omega =
\begin{pmatrix}
\omega_1 & \omega_4 & \omega_6 \\
\omega_4 & \omega_2 & \omega_5 \\
\omega_6 & \omega_5 & \omega_3 \\
\end{pmatrix}.
\end{align}
The complex coordinate on $T^6$, $\vec{z}$, obeys the identifications on $\mathbb{C}^3$,
\begin{align}
\vec{z} \sim \vec{z} + \vec{e}_i \sim \vec{z} + \Omega \vec{e}_i, \quad i=1,2,3.
\end{align}
We introduce the background magnetic fluxes on $T^6$ \cite{Antoniadis:2009bg}.
It is written by
\begin{align}
F = \frac{1}{2} p_{x^ix^j} dx^i \wedge dx^j + \frac{1}{2} p_{y^iy^j} dy^i \wedge dy^j + p_{x^iy^j} dx^i \wedge dy^j,
\end{align}
where $x$ and $y$ are indirectly defined through $z$ and $\bar{z}$ as
\begin{align}
z^i = x^i + \Omega^i_j y^j, \quad \bar{z}^i = x^i + \bar{\Omega}^i_j y^j.
\end{align}
In terms of $z$ and $\bar{z}$, it is rewritten as
\begin{align}
F = \frac{1}{2} p_{z^iz^j} dz^i \wedge dz^j + \frac{1}{2} p_{\bar{z}^i\bar{z}^j} d\bar{z}^i \wedge d\bar{z}^j + p_{z^i\bar{z}^j} (idz^i \wedge d\bar{z}^j),
\end{align}
where
\begin{align}
&p_{z^iz^j} = (\bar{\Omega}-\Omega)^{-1} (\bar{\Omega}p_{xx}\bar{\Omega}+p_{yy}+p_{xy}^T\bar{\Omega}-\bar{\Omega}p_{xy}) (\bar{\Omega}-\Omega)^{-1}, \\
&p_{\bar{z}^i\bar{z}^j} = (\bar{\Omega}-\Omega)^{-1} (\Omega p_{xx}\Omega+p_{yy}+p_{xy}^T\Omega-\Omega p_{xy}) (\bar{\Omega}-\Omega)^{-1}, \\
&p_{z^i\bar{z}^j} = i(\bar{\Omega}-\Omega)^{-1} (\bar{\Omega}p_{xx}\Omega+p_{yy}+p_{xy}^T\Omega-\bar{\Omega}p_{xy}) (\bar{\Omega}-\Omega)^{-1}.
\end{align}
Preserving supersymmetry requires that the fluxes must be a $(1,1)$-form, hence, $F_{zz}=F_{\bar{z}\bar{z}}=0$.
In addition, to make our analysis simple we assume $p_{xx}=p_{yy}=0$.
They lead to the $F$-term condition,
\begin{align}
p_{xy}^T\Omega = \Omega p_{xy}.
\end{align}
Due to the Dirac quantization condition, the values in $p_{xy}$ are constrained by
\begin{align}
p_{xy} = 2\pi N^T,
\end{align}
where $N$ must be a $3\times 3$ integer matrix.
Here and hereafter, we assume symmetric $N$ matrix, $N=N^T$, which is required by $S$-symmetry as we will see in the following section.
Eventually, one can obtain the $F$-term condition,
\begin{align}
N\Omega = \Omega N,
\end{align}
and the fluxes,
\begin{align}
F = \pi (N(\textrm{Im}\Omega)^{-1})_{ij} (idz^i \wedge d\bar{z}^j).
\end{align}
The vector potential is led by $F=dA$ as
\begin{align}
A =& \pi \textrm{Im} \left(N(\vec{\bar{z}}+\vec{\bar{\zeta}})(\textrm{Im}\Omega)^{-1} d\vec{z}\right) \notag \\
=& -\frac{i\pi}{2}\left(N(\vec{\bar{z}}+\vec{\bar{\zeta}})(\textrm{Im}\Omega)^{-1}\right)dz^i
+ \frac{i\pi}{2}\left(N(\vec{z}+\vec{\zeta})(\textrm{Im}\Omega)^{-1}\right)d\bar{z}^i \\
\equiv& A_{z^i}dz^i + A_{\bar{z}^i}d\bar{z}^i, \notag
\end{align}
where $\vec{\zeta}$ is the Wilson line.
In this paper we assume the vanishing Wilson line, $\vec{\zeta} = 0$.

Now we are ready to derive the zero-modes on $T^6$ with background magnetic fluxes.
We consider fermion massless modes (zero-modes) of the spinor,
\begin{align}
\Psi(\vec{z},\vec{\bar{z}}) = (
\psi_{(+,+,+)},
\psi_{(+,+,-)},
\psi_{(+,-,+)},
\psi_{(+,-,-)},
\psi_{(-,+,+)},
\psi_{(-,+,-)},
\psi_{(-,-,+)},
\psi_{(-,-,-)}
)^T,
\end{align}
where $(\pm,\pm,\pm)$ denotes the chiralities for each complex plane on $T^6$.
They satisfy the massless Dirac equation,
\begin{align}
i \slash{D} \Psi(\vec{z}, \vec{\bar{z}}) &= \frac{i}{\pi R}
\begin{pmatrix}
0 & D_{z^3} & D_{z^2} & 0 & D_{z^1} & 0 & 0 & 0 \\
\bar{D}_{\bar{z}^3} & 0 & 0 & -D_{z^2} & 0 & -D_{z^1} & 0 & 0 \\
\bar{D}_{\bar{z}^2} & 0 & 0 & D_{z^3} & 0 & 0 & -D_{z^1} & 0 \\
0 & -\bar{D}_{\bar{z}^2} & \bar{D}_{\bar{z}^3} & 0 & 0 & 0 & 0 & D_{z^1} \\
\bar{D}_{\bar{z}^1} & 0 & 0 & 0 & 0 & D_{z^3} & D_{z^2} & 0 \\
0 & -\bar{D}_{\bar{z}^1} & 0 & 0 & \bar{D}_{\bar{z}^3} & 0 & 0 & -D_{z^2} \\
0 & 0 & -\bar{D}_{\bar{z}^1} & 0 & \bar{D}_{\bar{z}^2} & 0 & 0 & D_{z^3} \\
0 & 0 & 0 & \bar{D}_{\bar{z}^1} & 0 & -\bar{D}_{\bar{z}^2} & \bar{D}_{\bar{z}^3} & 0 \\
\end{pmatrix}
\begin{pmatrix}
\psi_{(+,+,+)} \\
\psi_{(+,+,-)} \\
\psi_{(+,-,+)} \\
\psi_{(+,-,-)} \\
\psi_{(-,+,+)} \\
\psi_{(-,+,-)} \\
\psi_{(-,-,+)} \\
\psi_{(-,-,-)} \\
\end{pmatrix} = 0,
\end{align}
where
\begin{align}
D_{z^i} = \partial_{z^i} -iA_{z^i}, \quad \bar{D}_{\bar{z}^i} = \bar{\partial}_{\bar{z}^i} -iA_{\bar{z}^i}.
\end{align}
$\Psi(\vec{z},\vec{\bar{z}})$ must obey the boundary conditions on $T^6$ given by
\begin{align}
\Psi(\vec{z}+\vec{e}_k) = e^{i\xi_{\vec{e}_k}(\vec{z})} \Psi(\vec{z}), \quad
\Psi(\vec{z}+\Omega\vec{e}_k) = e^{i\xi_{\Omega\vec{e}_k}(\vec{z})} \Psi(\vec{z}),
\end{align}
where $k=1,2,3$ and
\begin{align}
\xi_{\vec{e}_k}(\vec{z}) = \pi(N(\textrm{Im}\Omega)^{-1}\textrm{Im}\vec{z})_k, \quad
\xi_{\Omega\vec{e}_k}(\vec{z}) = \pi\textrm{Im}(N\bar{\Omega}(\textrm{Im}\Omega)^{-1}\vec{z})_k.
\end{align}
We concentrate on the case that only the spinor component with chiralities $(+,+,+)$, $\psi_{(+,+,+)}$, do not vanish.
Then $\psi_{(+,+,+)}$ obeys the Dirac equation,
\begin{align}
\bar{D}_{\bar{z}^i} \psi_{(+,+,+)} = 0.
\end{align}
The solution to this equation is given by \cite{Cremades:2004wa}
\begin{align}
\psi_N^{\vec{j}}(\vec{z},\Omega) = {\cal N} e^{i\pi (N\vec{z})^T(\textrm{Im}\Omega)^{-1}\textrm{Im}\vec{z}} \cdot \vartheta
\begin{bmatrix}
\vec{j}N^{-1} \\ 0 \\
\end{bmatrix}
(N\vec{z},N\Omega), \label{eq;zero-modes}
\end{align}
where the Riemann-theta function with characteristics, $\vartheta\begin{bmatrix}
\vec{j}N^{-1} \\ 0 \\
\end{bmatrix}
(N\vec{z},N\Omega)$, is defined by
\begin{align}
\vartheta
\begin{bmatrix}
\vec{a} \\ \vec{b}
\end{bmatrix}(\vec{z},\Omega') =
\sum_{\vec{m}\in\mathbb{Z}^3}
e^{\pi i(\vec{m}+\vec{a})^T\Omega'(\vec{m}+\vec{a})} e^{2\pi i(\vec{m}+\vec{a})^T(\vec{z}+\vec{b})}, \quad \Omega'\in {\cal H}_3, \quad \vec{a},\vec{b}\in\mathbb{R}^3.
\end{align}
The indices of three components $\vec{j}\in\mathbb{Z}^3$ label the degeneracy of zero-modes.
Note that zero-modes have the periodicity,
\begin{align}
\psi_N^{\vec{j}+N\vec{e}_k} = \psi_N^{\vec{j}}, \quad k=1,2,3.
\end{align}
Therefore the independent $\vec{j}$ are given by $|\det N|$ number of lattice points in $\Lambda_N$ which is spanned by basis vectors $N\vec{e}_k$, $k=1,2,3$.
We note that zero-modes in Eq.~(\ref{eq;zero-modes}) are well-defined only if $\Omega'=N\Omega$ is an element of the Siegel upper-half plane ${\cal H}_3$ defined as \cite{Siegel:1943}
\begin{align}
{\cal H}_3 = \{\Omega'\in GL(3,\mathbb{C})|{\Omega'}^T = \Omega',~\textrm{Im}\Omega'>0 \}.
\end{align}
The normalization condition of zero-modes can be taken as
\begin{align}
\int_{T^6} d^3zd^3\bar{z} \psi_N^{\vec{j}} (\psi_N^{\vec{k}})^* = (2^3 \det (\textrm{Im}\Omega))^{-1/2} \delta_{\vec{j},\vec{k}}.
\end{align}


\section{$Sp(6,\mathbb{Z})$ modular symmetry}
\label{sec:Sp6Z}

In this section, we study $Sp(6,\mathbb{Z})$ modular symmetry.
$Sp(6,\mathbb{Z})$ is a symplectic group with the dimension 3 and over the integers $\mathbb{Z}$.
We will show that zero-modes on the magnetized $T^6$ behave as Siegel modular forms for subgroups of $Sp(6,\mathbb{Z})$.


\subsection{Modular transformation and Siegel modular forms}

Firstly we give a brief review of the modular symmetry on $T^6$ and Siegel modular forms \cite{Ding:2020zxw,Siegel:1943,Igusa:1972,Freitag:1983,Freitag:1991,Klingen:1990,Geer:2008,Fay:1973,Mumford:1984,Kikuchi:2023awe}.
The modular transformation $\gamma\in Sp(6,\mathbb{Z})$ is defined as the basis transformation of the lattice $\Lambda$ defining $T^6\simeq \mathbb{C}^3/\Lambda$.
It is given as $6\times 6$ integer matrices $\gamma$,
\begin{align}
\gamma =
\begin{pmatrix}
A & B \\
C & D \\
\end{pmatrix}
\in Sp(6,\mathbb{Z}),
\end{align}
satisfying
\begin{align}
\gamma J \gamma^T = J, \quad J=
\begin{pmatrix}
\bm{0}_3 & \bm{1}_3 \\
-\bm{1}_3 & \bm{0}_3 \\
\end{pmatrix}.
\end{align}
Under the modular transformation, the complex structure moduli of $T^6$, $\Omega$, are transformed as
\begin{align}
\Omega \rightarrow (A\Omega + B)(C\Omega + D)^{-1}. \label{eq:moduli_trans}
\end{align}
There are seven generators, $S$ and $T_i$ ($i=1,2,...,6$) defined by
\begin{align}
S = 
\begin{pmatrix}
\bm{0}_3 & \bm{1}_3 \\
-\bm{1}_3 & \bm{0}_3 \\
\end{pmatrix}, \quad
T_i =
\begin{pmatrix}
\bm{1}_3 & B_i \\
\bm{0}_3 & \bm{1}_3 \\
\end{pmatrix},
\quad i = 1,2,...,6,
\end{align}
where
\begin{align}
&B_1 =
\begin{pmatrix}
1 & 0 & 0 \\
0 & 0 & 0 \\
0 & 0 & 0 \\
\end{pmatrix}, \quad
B_2 =
\begin{pmatrix}
0 & 0 & 0 \\
0 & 1 & 0 \\
0 & 0 & 0 \\
\end{pmatrix}, \quad
B_3 =
\begin{pmatrix}
0 & 0 & 0 \\
0 & 0 & 0 \\
0 & 0 & 1 \\
\end{pmatrix}, \\
&B_4 =
\begin{pmatrix}
0 & 1 & 0 \\
1 & 0 & 0 \\
0 & 0 & 0 \\
\end{pmatrix}, \quad
B_5 =
\begin{pmatrix}
0 & 0 & 0 \\
0 & 0 & 1 \\
0 & 1 & 0 \\
\end{pmatrix}, \quad
B_6 =
\begin{pmatrix}
0 & 0 & 1 \\
0 & 0 & 0 \\
1 & 0 & 0 \\
\end{pmatrix}.
\end{align}
Thus the complex structure moduli $\Omega$ have $S$-symmetry and six $T$-symmetries.

The Siegel modular forms are holomorphic functions of $\Omega$ transformed as
\begin{align}
f^j(\Omega) \rightarrow f^j(\gamma:\Omega) = [\det(C\Omega+D)]^{k} \rho(\gamma)^{j\ell} f^\ell(\Omega),
\end{align}
where $k$ is the so-called modular weight and $\rho$ is a unitary representation of subgroup of $Sp(6,\mathbb{Z})$.
Generally $\rho$ fulfills $\rho(h)=\mathbb{I}$ for elements $h\in G \subset Sp(6,\mathbb{Z})$ and forms a finite group.


\subsection{Modular symmetry in zero-modes}

Next we study the modular transformations for zero-modes on $T^6$ \footnote{Modular symmetry in magnetized $T^{2g}$ was classified in Ref.~\cite{Kikuchi:2023awe}.}.
Under the modular transformation, the complex coordinate on $T^6$, $\vec{z}$, is transformed as
\begin{align}
\vec{z} \rightarrow {(C\Omega+D)^{-1}}^T \vec{z}.
\end{align}
Using this and Eq.~(\ref{eq:moduli_trans}), we can show that zero-modes on $T^6$ which are the functions of $\vec{z}$ and $\Omega$ behave as the Siegel modular forms for subgroup of $Sp(6,\mathbb{Z})$.
Indeed, when $(N\Omega)^T=N\Omega$, $N=N^T$ and $\Omega=\Omega^T$ are fulfilled, $S$-transformation of zero-modes can be derived as
\begin{align}
S: \psi^{\vec{j}}_N(\vec{z},\Omega) &= \psi^{\vec{j}}_N(-\Omega^{-1}\vec{z}, -\Omega^{-1}) 
= \sqrt{\det (-\Omega)} \sum_{\vec{k}\in\Lambda_N}\rho(S)^{\vec{j}\vec{k}} \psi^{\vec{k}}_N(\vec{z},\Omega), 
\end{align}
where
\begin{align}
\rho(S)^{\vec{j}\vec{k}} = \frac{e^{3\pi i/4}}{\sqrt{\det N}} e^{2\pi i \vec{j}^T N^{-1} \vec{k}},
\quad \vec{j}, \vec{k}\in \Lambda_N. \label{eq:rho_S}
\end{align}
In this way, the $F$-term condition for symmetric $N$ and $\Omega$, $N\Omega=\Omega N$, is also the consistency condition of $S$-transformation.
When $(N\Omega)^T=N\Omega$, $(NB)^T=NB$ and $(NB)^{kk}\in2\mathbb{Z}$ for $k=1,2,3$ are fulfilled, $T_i$-transformation of zero-modes can be derived as
\begin{align}
T_i: \psi^{\vec{j}}_N(\vec{z},\Omega) &= \psi^{\vec{j}}_N(\vec{z}, \Omega+B_i)
= \sum_{\vec{k}\in\Lambda_N}\rho(T_i)^{\vec{j}\vec{k}} \psi^{\vec{k}}_N(\vec{z},\Omega),
\end{align}
where
\begin{align}
\rho(T_i)^{\vec{j}\vec{k}} = e^{\pi i \vec{j}^T N^{-1}B_i \vec{j}} \delta_{\vec{j},\vec{k}},
\quad \vec{j}, \vec{k}\in \Lambda_N. \label{eq:rho_T}
\end{align}
Thus zero-modes on $T^6$ have the modular symmetry as the geometrical symmetry.
Taking $\vec{z}$ to the modular invariant point $\vec{z}=0$, zero-modes on $T^6$ become exactly the Siegel modular forms for subgroup of $Sp(6,\mathbb{Z})$.
Note that which $T$-transformations are well-defined in zero-modes depend on the structure of $N$ matrix.
$T$-symmetries not satisfying $(NB)^T=NB$ and $(NB)^{kk}\in2\mathbb{Z}$ break in zero-modes.

Here and hereafter we concentrate on $N$ matrix possessing three different eigenvalues.
This means that $N$ matrix cannot be expanded by $\bm{1}_3$ and $N^{-1}$.
If $N = \alpha \bm{1}_3 + \beta N^{-1}$ consists, we obtain
\begin{align}
N^2 -\alpha N - \beta \bm{1}_3 = 0 \rightarrow 
\begin{pmatrix}
\lambda_1^2 \\ \lambda_2^2 \\ \lambda_3^2 \\
\end{pmatrix}
- \alpha
\begin{pmatrix}
\lambda_1 \\ \lambda_2 \\ \lambda_3 \\
\end{pmatrix}
-\beta
\begin{pmatrix}
1 \\ 1 \\ 1 \\
\end{pmatrix}
= 0,
\end{align}
where $\lambda_1$, $\lambda_2$ and $\lambda_3$ are eigenvalues of $N$.
This equation requires at least two degenerate eigenvalues.
Therefore $N = \alpha \bm{1}_3 + \beta N^{-1}$ does not consist for $N$ matrix with three different eigenvalues.

Since $N$ matrix possesses three different eigenvalues, there are three independent symmetric matrices commuting to $N$ matrix.
By  $N \neq \alpha \bm{1}_3 + \beta N^{-1}$, we find that  such three symmetric matrices are given by $\bm{1}_3$, $N$ and $N^{-1}$.
To satisfy the $F$-term condition $N\Omega=\Omega N$, $\Omega$ must be given by the linear combination of independent symmetric matrices commuting to $N$ matrix.
Thus $\Omega$ can be written as
\begin{align}
\Omega = k_1\bm{1}_3 + k_2N + k_3N^{-1},
\end{align}
where $k_i$ ($i=1,2,3$) are any complex values.
Similarly, the condition $NB=BN$ restricts the structure of integral symmetric matrix $B$ to
\begin{align}
B = n_1\bm{1}_3 + n_2 \frac{1}{\textrm{lcm}(N)} N + n_3 \frac{\det N}{\textrm{lcm}(N)^2}N^{-1},
\end{align}
where $\textrm{lcm}(N)$ are the least common multiple of $N$ matrix elements, and $n_i$ ($i=1,2,3$) are any real values but $B$ must be an integral matrix.
We note that $\frac{1}{\textrm{lcm}(N)} N$ and $\frac{\det N}{\textrm{lcm}(N)^2}N^{-1}$ are integral matrices.
Obviously there are three independent $B$ matrices, that is, three independent $T$-transformations.
Thus zero-modes with $N$ matrix possessing three different eigenvalues have $S$-symmetry and three $T$-symmetries.


\section{Examples of Siegel modular forms}
\label{sec:Examples_Siegel}

In this section, we show some examples of the Siegel modular forms which are constructed from zero-modes on $T^6$.
As we have mentioned in the previous section, we use $N$ matrix with three different eigenvalues where three independent $T$-transformations exist in zero-modes.
We show five examples at $\det N =2,3,4,5$ and 6.


\subsection{$\det N=2$}

Let us choose the following $N$ matrix,
\begin{align}
N =
\begin{pmatrix}
-2 & 1 & -1 \\
1 & -2 & 1 \\
-1 & 1 & 0 \\
\end{pmatrix}.
\end{align}
The determinant of $N$ is 2.
Therefore there are two independent degenerate zero-modes,
\begin{align}
\psi^{\vec{j}}_N(\vec{z},\Omega) =
\begin{pmatrix}
\psi_N^{\begin{psmallmatrix}0\\0\\0\\\end{psmallmatrix}},
\psi_N^{\begin{psmallmatrix}-1\\0\\0\\\end{psmallmatrix}}
\end{pmatrix}.
\end{align}
These zero-modes are even modes under $\vec{z}\to -\vec{z}$ and do not vanish at $\vec{z}=0$.
Thus, even modes at $\vec{z}=0$,
\begin{align}
\label{eq:N=2}
\hat{\psi}^{\vec{j}}_{\bm{2}}(\Omega) =
\begin{pmatrix}
\hat{\psi}^0_{\bm{2}}(\Omega) \\
\hat{\psi}^1_{\bm{2}}(\Omega) \\
\end{pmatrix}
\equiv
\begin{pmatrix}
\psi_N^{\begin{psmallmatrix}0\\0\\0\\\end{psmallmatrix}}(0,\Omega) \\
\psi_N^{\begin{psmallmatrix}-1\\0\\0\\\end{psmallmatrix}}(0,\Omega) \\
\end{pmatrix},
\end{align}
can be regarded as the Siegel modular forms of weight 1/2.
They are mapped into themselves under $S$-transformation and following three $T$-transformations,
\begin{align}
T_I =
\begin{pmatrix}
\bm{1}_3 & B_I \\
\bm{0}_3 & \bm{1}_3 \\
\end{pmatrix}, \quad
T_{II} =
\begin{pmatrix}
\bm{1}_3 & B_{II} \\
\bm{0}_3 & \bm{1}_3 \\
\end{pmatrix}, \quad
T_{III} =
\begin{pmatrix}
\bm{1}_3 & B_{III} \\
\bm{0}_3 & \bm{1}_3 \\
\end{pmatrix},
\end{align}
where
\begin{align}
&B_I = \bm{1}_3 =
\begin{pmatrix}
1 & 0 & 0 \\
0 & 1 & 0 \\
0 & 0 & 1 \\
\end{pmatrix}, \\
&B_{II} = 2\bm{1}_3 + N =
\begin{pmatrix}
0 & 1 & -1 \\
1 & 0 & 1 \\
-1 & 1 & 2 \\
\end{pmatrix}, \\
&B_{III} = -\bm{1}_3 -2N^{-1} =
\begin{pmatrix}
0 & 1 & 1 \\
1 & 0 & -1 \\
1 & -1 & -4 \\
\end{pmatrix}.
\end{align}
These $B$ matrices satisfy consistency conditions for $T$-transformations.
The unitary representation matrices of $S$, $T_I$, $T_{II}$ and $T_{III}$-transformations, $\rho(S)$, $\rho(T_I)$, $\rho(T_{II})$ and $\rho(T_{III})$, form a group, $\widetilde{S}_4 \simeq T'\rtimes Z_4$, whose order is 96 \footnote{The group isomorphism $\widetilde{S}_4 \simeq T'\rtimes Z_4$ is pointed out in Refs.~\cite{Liu:2020msy,Uchida:2023yaj}}.
Then, $\hat{\psi}^{\vec{j}}_{\bm{2}}(\Omega) $ in Eq.~(\ref{eq:N=2}) are a doublet under $\widetilde{S}_4$.


\subsection{$\det N=3$}

Let us choose the following $N$ matrix,
\begin{align}
N = 
\begin{pmatrix}
0 & 3 & 4 \\
3 & 0 & -4 \\
4 & -4 & -11 \\
\end{pmatrix}.
\end{align}
The determinant of $N$ is 3.
Therefore there are three independent degenerate zero-modes,
\begin{align}
\psi^{\vec{j}}_N(\vec{z},\Omega) =
\begin{pmatrix}
\psi_N^{\begin{psmallmatrix}2\\2\\0\\\end{psmallmatrix}},
\psi_N^{\begin{psmallmatrix}1\\1\\0\\\end{psmallmatrix}},
\psi_N^{\begin{psmallmatrix}0\\0\\0\\\end{psmallmatrix}}
\end{pmatrix}.
\end{align}
These zero-modes are decomposed into even and odd modes under $\vec{z}\to-\vec{z}$.
At $\vec{z}=0$, only even modes do not vanish.
Thus, even modes at $\vec{z}=0$,
\begin{align}
\label{eq:N=3}
\hat{\psi}^{\vec{j}}_{\bm{2}}(\Omega) =
\begin{pmatrix}
\hat{\psi}^0_{\bm{2}}(\Omega) \\
\hat{\psi}^1_{\bm{2}}(\Omega) \\
\end{pmatrix}
\equiv
\begin{pmatrix}
\frac{1}{\sqrt{2}}\begin{pmatrix}\psi_N^{\begin{psmallmatrix}2\\2\\0\\\end{psmallmatrix}}(0,\Omega) + \psi_N^{\begin{psmallmatrix}1\\1\\0\\\end{psmallmatrix}}(0,\Omega)\end{pmatrix} \\
\psi_N^{\begin{psmallmatrix}0\\0\\0\\\end{psmallmatrix}}(0,\Omega) \\
\end{pmatrix},
\end{align}
can be regarded as the Siegel modular forms of weight 1/2.
They are mapped into themselves under $S$-transformation and following three $T$-transformations,
\begin{align}
T_I =
\begin{pmatrix}
\bm{1}_3 & B_I \\
\bm{0}_3 & \bm{1}_3 \\
\end{pmatrix}, \quad
T_{II} =
\begin{pmatrix}
\bm{1}_3 & B_{II} \\
\bm{0}_3 & \bm{1}_3 \\
\end{pmatrix}, \quad
T_{III} =
\begin{pmatrix}
\bm{1}_3 & B_{III} \\
\bm{0}_3 & \bm{1}_3 \\
\end{pmatrix},
\end{align}
where
\begin{align}
&B_I = 2\bm{1}_3 =
\begin{pmatrix}
2 & 0 & 0 \\
0 & 2 & 0 \\
0 & 0 & 2 \\
\end{pmatrix}, \\
&B_{II} = \frac{16}{13}\bm{1}_3 + \frac{3}{13}N + \frac{3}{13}N^{-1} =
\begin{pmatrix}
0 & 2 & 0 \\
2 & 0 & 0 \\
0 & 0 & -2 \\
\end{pmatrix}, \\
&B_{III} = \frac{6}{13}\bm{1}_3 - \frac{17}{104}N + \frac{9}{104}N^{-1} =
\begin{pmatrix}
0 & 0 & -1 \\
0 & 0 & 1 \\
-1 & 1 & 2 \\
\end{pmatrix}.
\end{align}
These $B$ matrices satisfy consistency conditions for $T$-transformations.
The unitary representation matrices of $S$, $T_I$, $T_{II}$ and $T_{III}$-transformations, $\rho(S)$, $\rho(T_I)$, $\rho(T_{II})$ and $\rho(T_{III})$, form a group, $(Z_8 \times Z_2) \rtimes Z_2 \rtimes Z_3$, whose order is 96.
Then, $\hat{\psi}^{\vec{j}}_{\bm{2}}(\Omega) $ in Eq.~(\ref{eq:N=3}) are a doublet of this group.


\subsection{$\det N=4$}

Let us choose the following $N$ matrix,
\begin{align}
N = 
\begin{pmatrix}
-1&1&-1\\
1&-1&-1\\
-1&-1&2\\
\end{pmatrix}. \label{eq:N_matrix}
\end{align}
The determinant of $N$ is 4.
Therefore there are four independent degenerate zero-modes,
\begin{align}
\psi^{\vec{j}}_N(\vec{z},\Omega) =
\begin{pmatrix}
\psi_N^{\begin{psmallmatrix}0\\-1\\0\\\end{psmallmatrix}},
\psi_N^{\begin{psmallmatrix}0\\0\\0\\\end{psmallmatrix}},
\psi_N^{\begin{psmallmatrix}0\\0\\-1\\\end{psmallmatrix}},
\psi_N^{\begin{psmallmatrix}-1\\0\\0\\\end{psmallmatrix}}
\end{pmatrix}.
\end{align}
These zero-modes are decomposed into even and odd modes under $\vec{z}\to-\vec{z}$.
At $\vec{z}=0$, only even modes do not vanish.
Thus, even modes at $\vec{z}=0$,
\begin{align}
\hat{\psi}^{\vec{j}}_{\bm{3}}(\Omega) =
\begin{pmatrix}
\hat{\psi}^0_{\bm{3}}(\Omega) \\
\hat{\psi}^1_{\bm{3}}(\Omega) \\
\hat{\psi}^2_{\bm{3}}(\Omega) \\
\end{pmatrix}
\equiv
\begin{pmatrix}
\frac{1}{\sqrt{2}}\psi_N^{\begin{psmallmatrix}0\\-1\\0\\\end{psmallmatrix}}(0,\Omega)
+ \frac{1}{\sqrt{2}} \psi_N^{\begin{psmallmatrix}-1\\0\\0\\\end{psmallmatrix}}(0,\Omega) \\
\psi_N^{\begin{psmallmatrix}0\\0\\0\\\end{psmallmatrix}}(0,\Omega) \\
-\psi_N^{\begin{psmallmatrix}0\\0\\-1\\\end{psmallmatrix}}(0,\Omega) \\
\end{pmatrix}, \label{eq:modular_form_1/2_Delta96}
\end{align}
can be regarded as the Siegel modular forms of weight 1/2.
They are mapped into themselves under $S$-transformation and following three $T$-transformations,
\begin{align}
T_I =
\begin{pmatrix}
\bm{1}_3 & B_I \\
\bm{0}_3 & \bm{1}_3 \\
\end{pmatrix}, \quad
T_{II} =
\begin{pmatrix}
\bm{1}_3 & B_{II} \\
\bm{0}_3 & \bm{1}_3 \\
\end{pmatrix}, \quad
T_{III} =
\begin{pmatrix}
\bm{1}_3 & B_{III} \\
\bm{0}_3 & \bm{1}_3 \\
\end{pmatrix},
\end{align}
where
\begin{align}
&B_I = -\frac{4}{3}\bm{1}_3 + \frac{2}{3}N -\frac{4}{3}N^{-1} =
\begin{pmatrix}
-1 & 1 & 0 \\
1 & -1 & 0 \\
0 & 0 & 0 \\
\end{pmatrix}, \label{eq:BI} \\
&B_{II} = -\bm{1}_3 + N =
\begin{pmatrix}
-2 & 1 & -1 \\
1 & -2 & -1 \\
-1 & -1 & 1 \\
\end{pmatrix}, \label{eq:BII} \\
&B_{III} = -\frac{2}{3}\bm{1}_3-\frac{2}{3}N+\frac{4}{3}N^{-1} =
\begin{pmatrix}
-1 & -1 & 0 \\
-1 & -1 & 0 \\
0 & 0 & -2 \\
\end{pmatrix}. \label{eq:BIII}
\end{align}
These $B$ matrices satisfy consistency conditions for $T$-transformations.
The unitary representation matrices of $S$, $T_I$, $T_{II}$ and $T_{III}$-transformations, $\rho(S)$, $\rho(T_I)$, $\rho(T_{II})$ and $\rho(T_{III})$, form a group, $\widetilde{\Delta}(96)\simeq \Delta(48)\rtimes Z_8$, whose order is 384.
Then, $\hat{\psi}^{\vec{j}}_{\bm{3}}(\Omega) $ in Eq.~(\ref{eq:modular_form_1/2_Delta96}) correspond to a triplet under $\widetilde{\Delta}(96)\simeq \Delta(48)\rtimes Z_8$.


\subsection{$\det N=5$}

Let us choose the following $N$ matrix,
\begin{align}
N = 
\begin{pmatrix}
-3 & -2 &  2 \\
-2 & -3 & -2 \\
 2 & -2 & -7 \\
\end{pmatrix}.
\end{align}
The determinant of $N$ is 5.
Therefore there are five independent degenerate zero-modes,
\begin{align}
\psi^{\vec{j}}_N(\vec{z},\Omega) =
\begin{pmatrix}
\psi_N^{\begin{psmallmatrix}0\\0\\0\\\end{psmallmatrix}},
\psi_N^{\begin{psmallmatrix}-1\\-1\\0\\\end{psmallmatrix}},
\psi_N^{\begin{psmallmatrix}-2\\-2\\0\\\end{psmallmatrix}},
\psi_N^{\begin{psmallmatrix}-3\\-3\\0\\\end{psmallmatrix}},
\psi_N^{\begin{psmallmatrix}-4\\-4\\0\\\end{psmallmatrix}}
\end{pmatrix}.
\end{align}
These zero-modes are decomposed into even and odd modes under $\vec{z}\to-\vec{z}$.
At $\vec{z}=0$, only even modes do not vanish.
Thus, even modes at $\vec{z}=0$,
\begin{align}
\label{eq:N=5}
\hat{\psi}^{\vec{j}}_{\bm{3}}(\Omega) =
\begin{pmatrix}
\hat{\psi}^0_{\bm{3}}(\Omega) \\
\hat{\psi}^1_{\bm{3}}(\Omega) \\
\hat{\psi}^2_{\bm{3}}(\Omega) \\
\end{pmatrix}
\equiv
\begin{pmatrix}
\frac{1}{\sqrt{2}}\begin{pmatrix}\psi_N^{\begin{psmallmatrix}-2\\-2\\0\\\end{psmallmatrix}}(0,\Omega) + \psi_N^{\begin{psmallmatrix}-3\\-3\\0\\\end{psmallmatrix}}(0,\Omega)\end{pmatrix} \\
\frac{1}{\sqrt{2}}\begin{pmatrix}\psi_N^{\begin{psmallmatrix}-1\\-1\\0\\\end{psmallmatrix}}(0,\Omega) + \psi_N^{\begin{psmallmatrix}-4\\-4\\0\\\end{psmallmatrix}}(0,\Omega)\end{pmatrix} \\
\psi_N^{\begin{psmallmatrix}0\\0\\0\\\end{psmallmatrix}}(0,\Omega)
\end{pmatrix},
\end{align}
can be regarded as the Siegel modular forms of weight 1/2.
They are mapped into themselves under $S$-transformation and following three $T$-transformations,
\begin{align}
T_I =
\begin{pmatrix}
\bm{1}_3 & B_I \\
\bm{0}_3 & \bm{1}_3 \\
\end{pmatrix}, \quad
T_{II} =
\begin{pmatrix}
\bm{1}_3 & B_{II} \\
\bm{0}_3 & \bm{1}_3 \\
\end{pmatrix}, \quad
T_{III} =
\begin{pmatrix}
\bm{1}_3 & B_{III} \\
\bm{0}_3 & \bm{1}_3 \\
\end{pmatrix},
\end{align}
where
\begin{align}
&B_I = 2\bm{1}_3 =
\begin{pmatrix}
2 & 0 & 0 \\
0 & 2 & 0 \\
0 & 0 & 2 \\
\end{pmatrix}, \\
&B_{II} = -\frac{3}{2}\bm{1}_3 - \frac{1}{2}N =
\begin{pmatrix}
0 & 1 & -1 \\
1 & 0 & 1 \\
-1 & 1 & 2 \\
\end{pmatrix}, \\
&B_{III} = \frac{19}{2}\bm{1}_3 + \frac{7}{4}N - \frac{5}{4}N^{-1} =
\begin{pmatrix}
0 & 1 & 1 \\
1 & 0 & -1 \\
1 & -1 & -4 \\
\end{pmatrix}.
\end{align}
These $B$ matrices satisfy consistency conditions for $T$-transformations.
The unitary representation matrices of $S$, $T_I$, $T_{II}$ and $T_{III}$-transformations, $\rho(S)$, $\rho(T_I)$, $\rho(T_{II})$ and $\rho(T_{III})$, form a group, $A_5\times Z_8$, whose order is 480.
Then, $\hat{\psi}^{\vec{j}}_{\bm{3}}(\Omega)$ in Eq.~(\ref{eq:N=5}) correspond to a triplet of $A_5\times Z_8$.


\subsection{$\det N=6$}

Let us choose the following $N$ matrix,
\begin{align}
N = 
\begin{pmatrix}
-2 & -1 & -1 \\
-1 & -2 &  1 \\
-1 &  1 &  0 \\
\end{pmatrix}.
\end{align}
The determinant of $N$ is 6.
Therefore there are six independent degenerate zero-modes,
\begin{align}
\psi^{\vec{j}}_N(\vec{z},\Omega) =
\begin{pmatrix}
\psi_N^{\begin{psmallmatrix}0\\0\\0\\\end{psmallmatrix}},
\psi_N^{\begin{psmallmatrix}-1\\0\\0\\\end{psmallmatrix}},
\psi_N^{\begin{psmallmatrix}-1\\-1\\0\\\end{psmallmatrix}},
\psi_N^{\begin{psmallmatrix}-2\\-1\\0\\\end{psmallmatrix}},
\psi_N^{\begin{psmallmatrix}-2\\-2\\0\\\end{psmallmatrix}},
\psi_N^{\begin{psmallmatrix}-3\\-2\\0\\\end{psmallmatrix}}
\end{pmatrix}.
\end{align}
These zero-modes are decomposed into even and odd modes under $\vec{z}\to-\vec{z}$.
At $\vec{z}=0$, only even modes do not vanish.
Thus, even modes at $\vec{z}=0$,
\begin{align}
\label{eq:N=6}
\hat{\psi}^{\vec{j}}_{\bm{4}}(\Omega) =
\begin{pmatrix}
\hat{\psi}^0_{\bm{4}}(\Omega) \\
\hat{\psi}^1_{\bm{4}}(\Omega) \\
\hat{\psi}^2_{\bm{4}}(\Omega) \\
\hat{\psi}^3_{\bm{4}}(\Omega) \\
\end{pmatrix}
\equiv
\begin{pmatrix}
\frac{1}{\sqrt{2}}\begin{pmatrix}\psi_N^{\begin{psmallmatrix}-1\\-1\\0\\\end{psmallmatrix}}(0,\Omega) + \psi_N^{\begin{psmallmatrix}-2\\-2\\0\\\end{psmallmatrix}}(0,\Omega)\end{pmatrix} \\
\psi_N^{\begin{psmallmatrix}-2\\-1\\0\\\end{psmallmatrix}}(0,\Omega) \\
\frac{1}{\sqrt{2}}\begin{pmatrix}\psi_N^{\begin{psmallmatrix}-1\\0\\0\\\end{psmallmatrix}}(0,\Omega) + \psi_N^{\begin{psmallmatrix}-3\\-2\\0\\\end{psmallmatrix}}(0,\Omega)\end{pmatrix} \\
\psi_N^{\begin{psmallmatrix}0\\0\\0\\\end{psmallmatrix}}(0,\Omega)
\end{pmatrix},
\end{align}
can be regarded as the Siegel modular forms of weight 1/2.
They are mapped into themselves under $S$-transformation and following three $T$-transformations,
\begin{align}
T_I =
\begin{pmatrix}
\bm{1}_3 & B_I \\
\bm{0}_3 & \bm{1}_3 \\
\end{pmatrix}, \quad
T_{II} =
\begin{pmatrix}
\bm{1}_3 & B_{II} \\
\bm{0}_3 & \bm{1}_3 \\
\end{pmatrix}, \quad
T_{III} =
\begin{pmatrix}
\bm{1}_3 & B_{III} \\
\bm{0}_3 & \bm{1}_3 \\
\end{pmatrix},
\end{align}
where
\begin{align}
&B_I = \bm{1}_3 =
\begin{pmatrix}
1 & 0 & 0 \\
0 & 1 & 0 \\
0 & 0 & 1 \\
\end{pmatrix}, \\
&B_{II} = -3\bm{1}_3 -2N+6N^{-1} =
\begin{pmatrix}
0 & 1 & -1 \\
1 & 0 & 1 \\
-1 & 1 & 0 \\
\end{pmatrix}, \\
&B_{III} = -2\bm{1}_3 - N =
\begin{pmatrix}
0 & 1 & 1 \\
1 & 0 & -1 \\
1 & -1 & -2 \\
\end{pmatrix}.
\end{align}
These $B$ matrices satisfy consistency conditions for $T$-transformations.
The unitary representation matrices of $S$, $T_I$, $T_{II}$ and $T_{III}$-transformations, $\rho(S)$, $\rho(T_I)$, $\rho(T_{II})$ and $\rho(T_{III})$, form a group, $\widetilde{S}_4\rtimes A_4$, whose order is 1152.
Then, $\hat{\psi}^{\vec{j}}_{\bm{4}}(\Omega) $ in Eq.~(\ref{eq:N=6}) correspond to a quartet of $\widetilde{S}_4\rtimes A_4$.

In Table \ref{tab:examples_SiegelModularForms}, we summarize the results.
\begin{table}[H]
\centering
\renewcommand{\arraystretch}{1.5}
\begin{tabular}{c|c|c|c} \hline
$N$ & $\det N$ & $B_I,B_{II},B_{III}$ & Groups \\ \hline
$\begin{psmallmatrix}-2&1&-1\\1&-2&1\\-1&1&0\\\end{psmallmatrix}$
& 2 & $
\begin{psmallmatrix}
1 & 0 & 0 \\
0 & 1 & 0 \\
0 & 0 & 1 \\
\end{psmallmatrix},
\begin{psmallmatrix}
0 & 1 & -1 \\
1 & 0 & 1 \\
-1 & 1 & 2 \\
\end{psmallmatrix},
\begin{psmallmatrix}
0 & 1 & 1 \\
1 & 0 & -1 \\
1 & -1 & -4 \\
\end{psmallmatrix}
$ & $\widetilde{S}_4$ \\
$\begin{psmallmatrix}0&3&4\\3&0&-4\\4&-4&-11\\\end{psmallmatrix}$
& 3 & $
\begin{psmallmatrix}
2 & 0 & 0 \\
0 & 2 & 0 \\
0 & 0 & 2 \\
\end{psmallmatrix},
\begin{psmallmatrix}
0 & 2 & 0 \\
2 & 0 & 0 \\
0 & 0 & -2 \\
\end{psmallmatrix},
\begin{psmallmatrix}
0 & 0 & -1 \\
0 & 0 & 1 \\
-1 & 1 & 2 \\
\end{psmallmatrix}
$ & $(Z_8 \times Z_2) \rtimes Z_2 \rtimes Z_3$ \\
$\begin{psmallmatrix}-1&1&-1\\1&-1&-1\\-1&-1&2\\\end{psmallmatrix}$
& 4 & $
\begin{psmallmatrix}
-1 & 1 & 0 \\
1 & -1 & 0 \\
0 & 0 & 0 \\
\end{psmallmatrix},
\begin{psmallmatrix}
-2 & 1 & -1 \\
1 & -2 & -1 \\
-1 & -1 & 1 \\
\end{psmallmatrix},
\begin{psmallmatrix}
-1 & -1 & 0 \\
-1 & -1 & 0 \\
0 & 0 & -2 \\
\end{psmallmatrix}
$ & $\widetilde{\Delta}(96)$ \\
$\begin{psmallmatrix}
-3 & -2 &  2 \\
-2 & -3 & -2 \\
 2 & -2 & -7 \\
\end{psmallmatrix}$ & 5 & 
$\begin{psmallmatrix}
2 & 0 & 0 \\
0 & 2 & 0 \\
0 & 0 & 2 \\
\end{psmallmatrix},
\begin{psmallmatrix}
0 & 1 & -1 \\
1 & 0 & 1 \\
-1 & 1 & 2 \\
\end{psmallmatrix},
\begin{psmallmatrix}
0 & 1 & 1 \\
1 & 0 & -1 \\
1 & -1 & -4 \\
\end{psmallmatrix}$
& $A_5\times Z_8$ \\
$\begin{psmallmatrix}-2 & -1 & -1 \\-1 & -2 &  1 \\-1 &  1 &  0 \\\end{psmallmatrix}$ & 6 &
$\begin{psmallmatrix}
1 & 0 & 0 \\
0 & 1 & 0 \\
0 & 0 & 1 \\
\end{psmallmatrix},
\begin{psmallmatrix}
0 & 1 & -1 \\
1 & 0 & 1 \\
-1 & 1 & 0 \\
\end{psmallmatrix},
\begin{psmallmatrix}
0 & 1 & 1 \\
1 & 0 & -1 \\
1 & -1 & -2 \\
\end{psmallmatrix}$
 & $\widetilde{S}_4\rtimes A_4$ \\ \hline
\end{tabular}
\caption{Examples of the Siegel modular forms of weight 1/2 obtained from the zero-modes on $T^6$.}
\label{tab:examples_SiegelModularForms}
\end{table}


\section{The Siegel modular forms for $\widetilde{\Delta}(96)$}
\label{sec:mf}

In the previous section, we show some examples of the Siegel modular forms of weight 1/2 obtained from the zero-modes on $T^6$.
In this section, we focus on the Siegel modular forms of weight 1/2 for $\widetilde{\Delta}(96)$ in Eq.~(\ref{eq:modular_form_1/2_Delta96}) and study them in details including higher weights.


\subsection{Weight 1/2}

Firstly, we redefine the Siegel modular forms of weight 1/2 for $\widetilde{\Delta}(96)$ in Eq.~(\ref{eq:modular_form_1/2_Delta96}) as
\begin{align}
\begin{pmatrix}
\zeta(\Omega) \\
\eta(\Omega) \\
\theta(\Omega) \\
\end{pmatrix}
\equiv
\begin{pmatrix}
\frac{1}{\sqrt{2}}\psi_N^{\begin{psmallmatrix}0\\-1\\0\\\end{psmallmatrix}}(0,\Omega)
+ \frac{1}{\sqrt{2}} \psi_N^{\begin{psmallmatrix}-1\\0\\0\\\end{psmallmatrix}}(0,\Omega) \\
\psi_N^{\begin{psmallmatrix}0\\0\\0\\\end{psmallmatrix}}(0,\Omega) \\
-\psi_N^{\begin{psmallmatrix}0\\0\\-1\\\end{psmallmatrix}}(0,\Omega) \\
\end{pmatrix}
=
\begin{pmatrix}
\sqrt{2}\psi_N^{\begin{psmallmatrix}0\\-1\\0\\\end{psmallmatrix}}(0,\Omega) \\
\psi_N^{\begin{psmallmatrix}0\\0\\0\\\end{psmallmatrix}}(0,\Omega) \\
-\psi_N^{\begin{psmallmatrix}0\\0\\-1\\\end{psmallmatrix}}(0,\Omega) \\
\end{pmatrix}, \label{eq:zeetth}
\end{align}
where the minus sign in the third row is merely the convention.
Furthermore, we take $B_I$, $B_{II}$ and $B_{III}$ given by Eqs.~(\ref{eq:BI}), (\ref{eq:BII}) and (\ref{eq:BIII}) into $N$-diagonal basis.
\begin{align}
O^T B_I O =
\begin{pmatrix}
0 & 0 & 0 \\
0 & -2 & 0 \\
0 & 0 & 0 \\
\end{pmatrix}, \quad
O^T B_{II} O =
\begin{pmatrix}
\sqrt{3} & 0 & 0 \\
0 & -3 & 0 \\
0 & 0 & -\sqrt{3} \\
\end{pmatrix}, \quad
O^T B_{III} O =
\begin{pmatrix}
-2 & 0 & 0 \\
0 & 0 & 0 \\
0 & 0 & -2 \\
\end{pmatrix},
\end{align}
where
\begin{align}
O =
\begin{pmatrix}
-\frac{1}{\sqrt{6+2\sqrt{3}}} & \frac{1}{\sqrt{2}} & \frac{1}{\sqrt{6-2\sqrt{3}}} \\
-\frac{1}{\sqrt{6+2\sqrt{3}}} & -\frac{1}{\sqrt{2}} & \frac{1}{\sqrt{6-2\sqrt{3}}} \\
\frac{1+\sqrt{3}}{\sqrt{6+2\sqrt{3}}} & 0 & -\frac{1-\sqrt{3}}{\sqrt{6-2\sqrt{3}}} \\
\end{pmatrix}, \label{eq:orthogonal}
\end{align}
and it diagonalizes $N$ as
\begin{align}
O^T N O =
\begin{pmatrix}
\sqrt{3}+1 & 0 & 0 \\
0 & -2 & 0 \\
0 & 0 & -\sqrt{3}+1 \\
\end{pmatrix}.
\end{align}
Obviously they commute with $N$.
Also it can be checked that they satisfy $(NB)^{ii}\in2\mathbb{Z}$ ($i=1,2,3$).
Hence, these $B$ matrices fulfill consistency conditions for $T$-transformations.

Under $S$, $T_I$, $T_{II}$ and $T_{III}$-transformations, the triplet Siegel modular form we have constructed in Eq.~(\ref{eq:zeetth}) is transformed as
\begin{align}
\begin{pmatrix}
\zeta \\ \eta \\ \theta \\
\end{pmatrix}
\xrightarrow{S}
\sqrt{\det (-\Omega)} \rho(S)
\begin{pmatrix}
\zeta \\ \eta \\ \theta \\
\end{pmatrix}, \quad
\begin{pmatrix}
\zeta \\ \eta \\ \theta \\
\end{pmatrix}
\xrightarrow{T_{I,II,III}}
\rho(T_{I,II,III})
\begin{pmatrix}
\zeta \\ \eta \\ \theta \\
\end{pmatrix},
\end{align}
where
\begin{align}
\begin{aligned}
&\rho(S) = e^{-7\pi i/4}
\begin{pmatrix}
0 & \frac{1}{\sqrt{2}}i & \frac{1}{\sqrt{2}}i \\
\frac{1}{\sqrt{2}}i & \frac{1}{2}i & -\frac{1}{2}i \\
\frac{1}{\sqrt{2}}i & -\frac{1}{2}i & \frac{1}{2}i \\
\end{pmatrix}, \quad
\rho(T_{I}) =
\begin{pmatrix}
i & 0 & 0 \\
0 & 1 & 0 \\
0 & 0 & 1 \\
\end{pmatrix}, \\
&\rho(T_{II}) =
\begin{pmatrix}
e^{7\pi i/4} & 0 & 0 \\
0 & 1 & 0 \\
0 & 0 & -1 \\
\end{pmatrix}, \quad
\rho(T_{III}) =
\begin{pmatrix}
-1 & 0 & 0 \\
0 & 1 & 0 \\
0 & 0 & 1 \\
\end{pmatrix}.
\end{aligned} \label{eq:rhos}
\end{align}
We can find $\rho(T_I)=\rho(T_{II}^{6})$ and $\rho(T_{III})=\rho(T_{II}^{4})$ although $T_I \neq T_{II}^6$ and $T_{III} \neq T_{II}^4$.
That is, we can regard $\rho(S)$ and $\rho(T_{II})$ as generators of unitary representations although $T_I$ and $T_{III}$ cannot be generated by $S$ and $T_{II}$.
They form a group $\widetilde{\Delta}(96)\simeq \Delta(48)\rtimes Z_8$.
Actually, $\rho(S)$ and $\rho(T_{II})$ satisfy the algebraic relations,
\begin{align}
\begin{aligned}
&\rho(S)^2=-i\bm{1}_3, ~(\rho(S)\rho(T_{II}))^3 = \rho(T_{II})^{8} = (\rho(S)^{-1}\rho(T_{II})^{-1}\rho(S)\rho(T_{II}))^3 = \bm{1}_3, \\
&\rho(S)^2\rho(T_{II}) = \rho(T_{II})\rho(S)^2.
\end{aligned}
\end{align}
They are the same as the algebraic relations in Eq.~(\ref{eq:algebra_S_T}) in Appendix A,  which correspond to ones of $\widetilde{\Delta}(96)$.
In appendix \ref{app:group_theory}, we summarize the group theory of $\widetilde{\Delta}(96)$.
The structures of $\rho(S)$ and $\rho(T_{II})$ mean $(\zeta,\eta,\theta)$ belongs to $\bm{3}'_7$.
Thus, $(\zeta,\eta,\theta)$ is a Siegel modular form of weight 1/2 for $\widetilde{\Delta}(96)$.

Next we study $q$-expansions of $(\zeta, \eta, \theta)$.
Let us parametrize $\Omega$ as
\begin{align}
\Omega 
&= \omega_1 B_{I} + \omega_2 B_{II} + \omega_3 B_{III} \\
&= O
\begin{pmatrix}
\sqrt{3}\omega_2-2\omega_3 & 0 & 0 \\
0 & -2\omega_1-3\omega_2 & 0 \\
0 & 0 & -\sqrt{3}\omega_2-2\omega_3 \\
\end{pmatrix} O^T. \label{eq:omega_para}
\end{align}
This is a general expression of $\Omega$ commuting to $N$ in Eq.~(\ref{eq:N_matrix}).
Under $T$-transformations, $(\omega_1,\omega_2,\omega_3)$ are transformed as
\begin{align}
\begin{aligned}
&(\omega_1, \omega_2, \omega_3) \xrightarrow{T_I} (\omega_1+1, \omega_2, \omega_3), \\
&(\omega_1, \omega_2, \omega_3) \xrightarrow{T_{II}} (\omega_1, \omega_2+1, \omega_3), \\
&(\omega_1, \omega_2, \omega_3) \xrightarrow{T_{III}} (\omega_1, \omega_2,\omega_3+1).
\end{aligned} \label{eq:omega123toT123}
\end{align}
Using this parametrization we expand $(\zeta, \eta, \theta)$ by $q_i\equiv e^{\pi i\omega_i/4}$ ($i=1,2,3$).
As we saw they are transformed as
\begin{align}
\begin{aligned}
&(\zeta, \eta, \theta) \xrightarrow{T_I} (e^{\frac{\pi i}{2}}\zeta, \eta, \theta), \\
&(\zeta, \eta, \theta) \xrightarrow{T_{II}} (e^{\frac{7\pi i}{4}}\zeta, \eta, e^{\pi i}\theta), \\
&(\zeta, \eta, \theta) \xrightarrow{T_{III}} (e^{\pi i}\zeta, \eta, \theta),
\end{aligned} \label{eq:T-charges}
\end{align}
under $T$-transformations.
This and Eq.~(\ref{eq:omega123toT123}) restrict $(\zeta, \eta, \theta)$ to the following structures, 
\begin{align}
&\zeta = q_1^2q_2^7q_3^4\sum_{\vec{m}\in\mathbb{Z}^3} c^1_{m_1m_2m_3} q_1^{8m_1} q_2^{8m_2} q_3^{8m_3}, \\
&\eta = \sum_{\vec{m}\in\mathbb{Z}^3} c^2_{m_1m_2m_3} q_1^{8m_1} q_2^{8m_2} q_3^{8m_3}, \\
&\theta = q_2^4\sum_{\vec{m}\in\mathbb{Z}^3} c^3_{m_1m_2m_3} q_1^{8m_3} q_2^{8m_1} q_3^{8m_2},
\end{align}
where $c^1$, $c^2$ and $c^3$ denote constant order 3 tensors.
Substituting $\Omega$ in Eq.~(\ref{eq:omega_para}) to the definitions of $(\zeta, \eta, \theta)$ in Eq.~(\ref{eq:zeetth}), we actually obtain $q$-expansions,
\begin{align}
\begin{pmatrix}
\zeta \\ \eta \\ \theta \\
\end{pmatrix}
=
\begin{pmatrix}
\sqrt{2}(2q_1^{2}q_2^{7}q_3^{4}
+2q_1^{2}q_2^{7}q_3^{-4}
+2q_1^{2}q_2^{15}q_3^{-12}
+2q_1^{2}q_2^{15}q_3^{12}
+2q_1^{18}q_2^{31}q_3^{-4}
+2q_1^{18}q_2^{31}q_3^{4} + \cdots) \\
1
+2q_2^{16}
+4q_1^{8}q_2^{16}
+2q_2^{16}q_3^{16}
+2q_2^{16}q_3^{-16}
+4q_1^{8}q_2^{24} + \cdots \\
-(2q_2^{4}
+2q_2^{12}
+2q_1^{8}q_2^{12}
+4q_1^{8}q_2^{28}
+4q_1^{8}q_2^{28}q_3^{16}
+4q_1^{8}q_2^{28}q_3^{-16} + \cdots) \\
\end{pmatrix}. \label{eq:q-ex_zethet}
\end{align}
It is remarkable that powers of $q_1$ and $q_2$ are not negative while one of $q_3$ can be negative in $(\zeta,\eta,\theta)$.
This is because diag$(NB_I)\geq 0$ and diag$(NB_{II}) \geq 0$ but diag$(N)$ includes a negative eigenvalue.
At two cusps $\omega_1= i\infty$ and $\omega_2= i\infty$ where $q_1$ and $q_2$ vanish respectively, we obtain
\begin{align}
\begin{array}{llll}
\omega_1=i\infty: &
\zeta = 0, & \eta = 1+2q_2^{16}+\cdots, & \theta = -2q_2^4-2q_2^{12}+\cdots, \\
\omega_2=i\infty: &
\zeta = 0, & \eta = 1, & \theta = 0.
\end{array}
\end{align}

Finally we study higher weight Siegel modular forms for $\widetilde{\Delta}(96)$.
Higher weight Siegel modular forms can be constructed by tensor products of $(\zeta,\eta,\theta)$.
We show the Siegel modular forms up to weight 5 in appendix \ref{app:modularforms}.
The dimension of the Siegel modular forms of weight $k$ is given by $_{2k+2}C_2 =\frac{1}{2}(2k+2)(2k+1)$.
This can be shown by a mathematical induction.
We introduce a vector $Y_n$ containing the products of $n$ number of $(\zeta,\eta,\theta)$:
\begin{align}
Y_n \equiv
\begin{pmatrix}
\zeta^n \\
\zeta^{n-1}\eta \\
\vdots \\
\zeta\eta\theta^{n-2} \\
\zeta\theta^{n-1} \\
\eta^n \\
\eta^{n-1}\theta \\
\vdots \\
\eta \theta^{n-1} \\
\theta^n \\
\end{pmatrix}
\begin{array}{l}
\left.
\begin{matrix}
\\ \\ \\ \\ \\
\end{matrix}
\right\} \textrm{Products with one or more $\zeta$} \\
\left.
\begin{matrix}
\\ \\ \\ \\ \\
\end{matrix}
\right\} \textrm{Products without $\zeta$} \\
\end{array}.
\end{align}
We assume that $Y_{n-1}$ spans $_{n+1}C_2=\frac{1}{2}(n+1)n$ dimensional spaces.
$Y_{n-1}$ contains $\frac{1}{2}(n+1)n$ number of products; therefore this assumption means all products of $n-1$ number of $(\zeta,\eta,\theta)$ are linearly independent each other.
Then we will show $Y_n$ spans $_{n+2}C_2=\frac{1}{2}(n+2)(n+1)$ dimensional spaces.

First let us consider the products with one or more $\zeta$ in $Y_n$.
They are given by $\zeta Y_{n-1}$.
Since the overall factor $\zeta$ does not change the dimension of $Y_{n-1}$, $\zeta Y_{n-1}$ should span $\frac{1}{2}(n+1)n$ dimensional spaces if $Y_{n-1}$ spans $\frac{1}{2}(n+1)n$ dimensional spaces.
That is, they satisfy
\begin{align}
\sum_{j=0,1,...,n-1} \sum_{k=0,1,...,j} c_{jk} \zeta^{n-j} \eta^{j-k} \theta^{k} = 0 \to c_{jk}=0. \label{eq:c_jk=0}
\end{align}

Second we investigate products without $\zeta$.
As we have seen above, the lowest orders of $q_2$ in $\eta$ and $\theta$ are $q_2^0$ and $q_2^4$, respectively.
This means
\begin{align}
\eta^n \neq \sum_{j=1,2,...,n} c_j\eta^{n-j} \theta^{j}, \quad \forall c_j \in \mathbb{C},
\end{align}
because the left-hand side contains $q_2^0$ while the right-hand side contains $q_2^{4}$ as the lowest order.
In a similar way, we find
\begin{align}
\eta^{n-1}\theta \neq \sum_{j=2,3,...,n} c_j\eta^{n-j} \theta^{j}, \quad \forall c_j \in \mathbb{C},
\end{align}
because the left-hand side contains $q_2^{4}$ while the right-hand side contains $q_2^{8}$ as the lowest order.
Repeating this procedure, we can show
\begin{align}
\sum_{j=0,1,...,n} c_j\eta^{n-j} \theta^{j} = 0 \to c_j = 0. \label{eq:c_j=0}
\end{align}

Now we are ready to discuss the dimension of $Y_n$.
Let us consider the following equation,
\begin{align}
\sum_{j=0,1,...,n-1} \sum_{k=0,1,...,j} c_{jk} \zeta^{n-j} \eta^{j-k} \theta^{k}
+
\sum_{j=0,1,...,n} c_j\eta^{n-j} \theta^{j} = 0.
\end{align}
At $\omega_1=i\infty$ ($\zeta=0$), the first term vanishes.
Hence this equation is devided to two equations,
\begin{align}
\sum_{j=0,1,...,n-1} \sum_{k=0,1,...,j} c_{jk} \zeta^{n-j} \eta^{j-k} \theta^{k} = 0, \quad
\sum_{j=0,1,...,n} c_j\eta^{n-j} \theta^{j} = 0,
\end{align}
and they consist only if $c_{jk}=0$ and $c_j=0$ as shown in Eqs.~(\ref{eq:c_jk=0}) and (\ref{eq:c_j=0}).
Therefore, we find
\begin{align}
\sum_{j=0,1,...,n-1} \sum_{k=0,1,...,j} c_{jk} \zeta^{n-j} \eta^{j-k} \theta^{k}
+
\sum_{j=0,1,...,n} c_j\eta^{n-j} \theta^{j} = 0 \to c_{jk}=0,~c_j=0.
\end{align}
Thus all products in $Y_n$ should be linearly independent each other and spans $_{n+2}C_2=\frac{1}{2}(n+2)(n+1)$ dimensional spaces if $Y_{n-1}$ spans $_{n+1}C_2=\frac{1}{2}(n+1)n$ dimensional spaces.
Since we have already known that $Y_1=(\zeta, \eta, \theta)$ spans three dimensional spaces, this is true for all $Y_n$ $(n\geq1)$.


\subsection{Residual symmetries}

$S$, $T_I$, $T_{II}$ and $T_{III}$-transformations form the following algebraic relations,
\begin{align}
S^4 =
(ST_I^{-1}T_{II})^{12} =
(ST_{I}^{-2}T_{II})^{12} =
\mathbb{I}. \label{eq:algebra_N}
\end{align}
Thus we can find three invariant moduli corresponding to these algebraic relations.
In addition there are three invariant moduli under $T_I$, $T_{II}$ and $T_{III}$-transformations.
In Table \ref{tab:residuals}, we summarize these invariant moduli.
Note that the structures of the moduli are restricted by $N\Omega=\Omega N$.
Hence, it is diagonalized by the orthogonal matrix $O$ in Eq.~(\ref{eq:orthogonal}) as $N$.
\begin{table}[H]
\centering
\begin{tabular}{ccc} \hline
$\gamma$ & $\gamma:\Omega$ & Invariant moduli \\ \hline
$S$ & $-\Omega^{-1}$ & 
$\Omega_{S}\equiv O\begin{pmatrix}e^{\pm\frac{\pi i}{2}}&0&0\\0&e^{\pm\frac{\pi i}{2}}&0\\0&0&e^{\pm\frac{\pi i}{2}}\\\end{pmatrix}O^T$ \\
$ST_I^{-1}T_{II}$ & $-(\Omega-B_I+B_{II})^{-1}$ & 
$\Omega_{ST_I^{-1}T_{II}}\equiv O\begin{pmatrix}e^{\pm\frac{\pi i}{6}}&0&0\\0&e^{\pm\frac{2\pi i}{3}}&0\\0&0&e^{\pm\frac{5\pi i}{6}}\\\end{pmatrix}O^T$ \\
$ST_{I}^{-2}T_{II}$ & $-(\Omega-2B_I+B_{II})^{-1}$ & 
$\Omega_{ST_I^{-2}T_{II}}\equiv O\begin{pmatrix}e^{\pm\frac{\pi i}{6}}&0&0\\0&e^{\pm\frac{\pi i}{3}}&0\\0&0&e^{\pm\frac{5\pi i}{6}}\\\end{pmatrix}O^T$ \\
$T_I$ & $\Omega+B_I$ & 
$\Omega_{T_I}\equiv i\infty B_I$ ($\omega_1=i\infty$) \\
$T_{II}$ & $\Omega+B_{II}$ & 
$\Omega_{T_{II}}\equiv i\infty B_{II}$ ($\omega_2=i\infty$) \\
$T_{III}$ & $\Omega+B_{III}$ & 
$\Omega_{T_{III}}\equiv i\infty B_{III}$ ($\omega_3=i\infty$) \\
\hline
\end{tabular}
\caption{Invariant moduli corresponding to the algebraic relations in Eq.~(\ref{eq:algebra_N}) and $T$-transformations.
$\pm$ in the third column means any double sign.}
\label{tab:residuals}
\end{table}


\section{Numerical example}
\label{sec:Num}

In this section, we study quark flavor models using the Siegel modular forms for $\widetilde{\Delta}(96)$.
All possible structures of fermion mass matrices with $\widetilde{\Delta}(96)$ symmetry are classified 
in Table \ref{tab:matrices}  of   Appendix \ref{app:modularforms}.
For simplicity, we focus on the Siegel modular forms belonging to $\widetilde{\Delta}(96)$ singlets in our model building.
There are eight singlets $\bm{1}_q$, $q=0,1,...,7$, in $\widetilde{\Delta}(96)$.
The fermion mass matrices in this case are shown in the first row of Table \ref{tab:matrices}.

We start to discuss how to obtain quark mass hierarchies.
Quarks have large mass hierarchies as shown in Table \ref{tab:quark_masses}.
\begin{table}[H]
    \begin{center}
    \renewcommand{\arraystretch}{1.3}
    \begin{tabular}{c|cccc} \hline
      & $\frac{m_u}{m_t}\times10^{6}$ & $\frac{m_c}{m_t}\times10^3$ & $\frac{m_d}{m_b}\times10^4$ & $\frac{m_s}{m_b}\times10^2$ \\ \hline
      GUT scale values & $5.39$ & $2.80$ & $9.21$ & $1.82$ \\
      $1\sigma$ errors & $\pm 1.68$ & $\pm 0.12$ & $\pm 1.02$ & $\pm 0.10$ \\ \hline
    \end{tabular}
  \end{center}
      \caption{Quark mass ratios at the GUT scale $2\times 10^{16}$ GeV with $\tan \beta=5$ \cite{Antusch:2013jca,Bjorkeroth:2015ora}.}
\label{tab:quark_masses}
\end{table}
To reproduce these large mass hierarchies, we use residual symmetry around the cusp,
\begin{align}
\Omega \sim \Omega_{T_I} = i\infty B_I.
\end{align}
When $\Omega$ lies on the vicinity of $\Omega_{T_I}$, the Siegel modular forms $f(\Omega)$ with the residual charge $r$ can be expanded by powers of $\omega_1$ as
\begin{align}
f(\Omega) \sim q_1^r, \quad q_1 \equiv e^{\pi i \omega_1/4}, \quad \omega_1 \sim i\infty. \label{eq:fsimq1}
\end{align}
Actually as in Eq.~(\ref{eq:q-ex_zethet}) $(\zeta, \eta, \theta)$ is expanded as
\begin{align}
\begin{pmatrix}
\zeta \\ \eta \\ \theta \\
\end{pmatrix}
\sim
\begin{pmatrix}
\sqrt{2} q_1^2 (2q_2^7q_3^4+2q_2^7q_3^{-4}+2q_2^{15}q_3^{12}+2q_2^{15}q_3^{-12}+\cdots) \\
1+2q_2^{16} + 2q_2^{16}q_3^{16}+\cdots \\
-2q_2^4-2q_2^{12}+\cdots \\
\end{pmatrix},
\end{align}
at $\Omega \sim i\infty B_I$ $(\omega_1\sim i\infty)$.
Since $(\zeta, \eta, \theta)$ has $T_I$-charge $(2, 0, 0)$ as shown in Eq.~(\ref{eq:T-charges}), this is consistent with Eq.~(\ref{eq:fsimq1}).
Thus in the vicinity of $\Omega_{T_I}$ the modular forms become hierarchical depending on their $T_I$-charges.
This behavior yields the possibility to reproduce large mass hierarchies of quarks.
To obtain more realistic mass hierarchies of quarks, we further assume $\omega_2\sim i\infty$ and $|\omega_2| < |\omega_1|$, where $|q_1|<|q_2|<1$.
Then $(\zeta, \eta, \theta)$ is simply evaluated as
\begin{align}
\begin{pmatrix}
\zeta \\ \eta \\ \theta \\
\end{pmatrix}
\sim
\begin{pmatrix}
4\sqrt{2} q_1^2 q_2^7 \\
1 \\
-2q_2^4 \\
\end{pmatrix}, \label{eq:zetaetatheta_q1q2}
\end{align}
where we have the hierarchy $\zeta,\theta \ll \eta\sim 1$.
The hierarchy between $\zeta$ and $\theta$ is controlled by the values of $\omega_1$ and $\omega_2$.

Next, let us see the behaviors of singlet Siegel modular forms in Eq.~(\ref{eq:zetaetatheta_q1q2}).
In Table \ref{tab:T-charges_of_singlets}, we show $T_I$, $T_{II}$ and $T_{III}$-charges of $\widetilde{\Delta}(96)$ singlets.
\begin{table}[H]
\centering
\begin{tabular}{c|cccccccc} \hline
& $\bm{1}_0$ & $\bm{1}_1$ & $\bm{1}_2$ & $\bm{1}_3$ & $\bm{1}_4$ & $\bm{1}_5$ & $\bm{1}_6$ & $\bm{1}_7$ \\ \hline
$T_I$-charges & 0 & 6 & 4 & 2 & 0 & 6 & 4 & 2 \\
$T_{II}$-charges & 0 & 1 & 2 & 3 & 4 & 5 & 6 & 7 \\
$T_{III}$-charges & 0 & 4 & 0 & 4 & 0 & 4 & 0 & 4 \\ \hline
\end{tabular}
\caption{$T_I$, $T_{II}$ and $T_{III}$-charges of $\widetilde{\Delta}(96)$ singlets.}
\label{tab:T-charges_of_singlets}
\end{table}
Up to weight 5, we can find eight modular forms belonging to singlets,
\begin{align}
Y^{(3/2)}_{\bm{1}_7}, ~Y^{(2)}_{\bm{1}_4}, ~Y^{(3)}_{\bm{1}_6}, ~Y^{(7/2)}_{\bm{1}_3}, ~Y^{(4)}_{\bm{1}_{0a}}, ~Y^{(4)}_{\bm{1}_{0b}}, ~Y^{(9/2)}_{\bm{1}_5}, ~Y^{(5)}_{\bm{1}_2}.
\end{align}
When $\zeta,\theta \ll \eta\sim 1$, they are evaluated as
\begin{align}
\begin{aligned}
&Y^{(3/2)}_{\bm{1}_7} \simeq
\frac{\sqrt{6}}{2}\zeta\simeq 4\sqrt{3}q_1^2q_2^7, \quad
Y^{(2)}_{\bm{1}_4} \simeq
\frac{2\sqrt{3}}{3}\theta\simeq -\frac{4\sqrt{3}}{3}q_2^4, \quad
Y^{(3)}_{\bm{1}_6} \simeq
\frac{3}{2}\zeta^{2} \simeq 48q_1^4q_2^{14}, \\
&Y^{(7/2)}_{\bm{1}_3} \simeq
\sqrt{2}\zeta\theta \simeq -16q_1^2q_2^{11}, \quad
Y^{(4)}_{\bm{1}_{0a}} \simeq
\frac{1}{4\sqrt{6}}, \quad
Y^{(4)}_{\bm{1}_{0b}} \simeq
\frac{4}{3}\theta^{2} \simeq \frac{16}{3}q_2^8, \\
&Y^{(9/2)}_{\bm{1}_5} \simeq
-\frac{3}{2\sqrt{6}}\zeta^{3} \simeq -64\sqrt{3}q_1^6q_2^{21}, \quad
Y^{(5)}_{\bm{1}_2} \simeq
\frac{1}{\sqrt{3}}\zeta^{2}\theta \simeq -\frac{64}{\sqrt{3}}q_1^4q_2^{18},
\end{aligned}
\end{align}
in the first order approximation of $\zeta$ and $\theta$.
In the following we ignore $Y^{(4)}_{\bm{1}_{0b}}$ because it is negligible comparing with $Y^{(4)}_{\bm{1}_{0a}}$ due to $\theta\ll 1$.

Now we are ready to build the quark flavor model reproducing large mass hierarchies in the vicinity of $\Omega_{T_I}$ using singlet Siegel modular forms.
We study the model with the assignments shown in Table \ref{eq:assignments}.
\begin{table}[H]
\centering
\begin{tabular}{c|ccccc} \hline
& $Q$ & $u_R$ & $d_R$ & $H_u$ & $H_d$ \\ \hline
Weights & $(-1/2, -5/2, -3/2)$ & $(0, -1/2, -5/2)$ & $(-5/2, -1, -5/2)$ & 0 & 0 \\
Irr. reps. & $(\bm{1}_2, \bm{1}_6, \bm{1}_0)$ & $(\bm{1}_1, \bm{1}_4, \bm{1}_0)$ & $(\bm{1}_0, \bm{1}_7, \bm{1}_0)$ & $\bm{1}_0$ & $\bm{1}_0$ \\ \hline
\end{tabular}
\caption{Assignments in our model.}
\label{eq:assignments}
\end{table}
Then up and down quark mass matrices $M_u$ and $M_d$ are given by
\begin{align}
M_u &= y_uv_u
\begin{pmatrix}
0 & 0 & C^{-1/2}\alpha^{13}Y_{\bm{1}_6}^{(3)} \\
0 &C^{-1/2}\alpha^{22}Y_{\bm{1}_6}^{(3)} & C^{1/2}\alpha^{23}Y_{\bm{1}_2}^{(5)} \\
C^{-5/4}\alpha^{31}Y_{\bm{1}_7}^{(3/2)} & C^{-1}\alpha^{32}Y_{\bm{1}_4}^{(2)} & \alpha^{33}Y_{\bm{1}_{0a}}^{(4)} \\
\end{pmatrix} \\
&\simeq y_uv_u
\begin{pmatrix}
0 & 0 & C^{-1/2}\alpha^{13}48q_1^4q_2^{14} \\
0 & C^{-1/2}\alpha^{22}48q_1^4q_2^{14} & -C^{1/2}\alpha^{23}\frac{64}{\sqrt{3}}q_1^4q_2^{18} \\
C^{-5/4}\alpha^{31}4\sqrt{3}q_1^2q_2^7 & -C^{-1}\alpha^{32}\frac{4\sqrt{3}}{3}q_2^4 & \alpha^{33}\frac{1}{4\sqrt{6}} \\
\end{pmatrix}, \label{eq:up_mass_mtx} \\
 M_d &=y_d v_d
\begin{pmatrix}
C^{-1/2}\beta^{11}Y_{\bm{1}_6}^{(3)} &C^{-5/4}\beta^{12}Y_{\bm{1}_7}^{(3/2)} & C^{-1/2}\beta^{13}Y_{\bm{1}_6}^{(3)} \\
C^{1/2}\beta^{21}Y_{\bm{1}_2}^{(5)} &C^{-1/4}\beta^{22}Y_{\bm{1}_3}^{(7/2)} & C^{1/2}\beta^{23}Y_{\bm{1}_2}^{(5)} \\
\beta^{31}Y_{\bm{1}_{0a}}^{(4)} & 0 &\beta^{33}Y_{\bm{1}_{0a}}^{(4)} \\
\end{pmatrix} \\
&\simeq y_d v_d
\begin{pmatrix}
C^{-1/2}\beta^{11}48q_1^4q_2^{14} & C^{-5/4}\beta^{12}4\sqrt{3}q_1^2q_2^7 & C^{-1/2}\beta^{13}48q_1^4q_2^{14} \\
-C^{1/2}\beta^{21}\frac{64}{\sqrt{3}}q_1^4q_2^{18} & -C^{-1/4}\beta^{22}16q_1^2q_2^{11} & -C^{1/2}\beta^{23}\frac{64}{\sqrt{3}}q_1^4q_2^{18} \\
\beta^{31}\frac{1}{4\sqrt{6}} & 0 &\beta^{33}\frac{1}{4\sqrt{6}} \\
\end{pmatrix}, \label{eq:down_mass_mtx}
\end{align}
where $v_u$ and $v_d$ denote vacuum expectation values of the up and down sector Higgs fields, respectively, and 
$y_u$ and $y_d$ are unknown overall coefficients.
In addition, we denote $C=(2^3\det\textrm{Im}\Omega)$ and 
its  powers  are originated from the ratio of Kahler metric.
We have ambiguities in relative normalizations of modular forms, which are written by 
$\alpha^{ij}$ and $\beta^{ij}$ in the above equations.
However, we do not expect that these normalization factors lead to a large hierarchy.
Thus, it would be natural that all of $\alpha^{ij}$ and $\beta^{ij}$ are  similar values, i.e. parameters of ${\cal O}(1)$.
Also we regard  values of moduli $(\omega_1, \omega_2,\omega_3)$ as free parameters.

To generate large mass hierarchies, first let us fix $\omega_1$ near the cusp, $\omega_1=1.3i\sim i\infty$, where $q_1=0.360$.
In addition, we fix $(\omega_2, \omega_3)$ at $(0.6i,0)$, where $q_2=0.624$ and $q_3=1$.
Hence, we choose the following point of moduli,
\begin{align}
(\omega_1, \omega_2, \omega_3) = (1.3i, 0.6i, 0),
\end{align}
where
\begin{align}
(q_1,q_2,q_3) = (0.360, 0.624, 1).
\end{align}
When we choose $\alpha^{ij}$ and $\beta^{ij}$ to be $\pm 1$ as
\begin{align}
&\begin{pmatrix}
\text{-} & \text{-} & \alpha^{13} \\
\text{-} & \alpha^{22} & \alpha^{23} \\
\alpha^{31} & \alpha^{32} & \alpha^{33} \\
\end{pmatrix}
=
\begin{pmatrix}
\text{-} & \text{-} & 1 \\
\text{-} & 1 & -1 \\
1 & -1 & 1 \\
\end{pmatrix}, \quad
\begin{pmatrix}
\beta^{11} & \beta^{12} & \beta^{13} \\
\beta^{21} & \beta^{22} & \beta^{23} \\
\beta^{31} & \text{-} & \beta^{33} \\
\end{pmatrix}
=
\begin{pmatrix}
1 & 1 & 1 \\
1 & -1 & -1 \\
1 & \text{-} & -1 \\
\end{pmatrix},
\end{align}
we obtain the quark mass ratios,
\begin{align}
&(m_u, m_c, m_t)/m_t = (1.08\times 10^{-5}, 1.06\times 10^{-3}, 1), \\
&(m_d, m_s, m_b)/m_b = (1.82\times 10^{-3}, 3.40\times 10^{-2}, 1),
\end{align}
absolute values of the CKM matrix elements,
\begin{align}
|V_{\textrm{CKM}}| =
\begin{pmatrix}
0.974 & 0.228 & 0.00441 \\
0.228 & 0.974 & 0.0163 \\
0.000581 & 0.0168 & 1.00 \\
\end{pmatrix},
\end{align}
and the value of Jarlskog invariant,
\begin{align}
J_{\textrm{CP}} = |\textrm{Im}(V_{\textrm{CKM}}^{us}V_{\textrm{CKM}}^{cb}(V_{\textrm{CKM}}^{ub}V_{\textrm{CKM}}^{cs})^*)| = 0.
\end{align}
Results are summarized in Table \ref{tab:fitting_one}.
\begin{table}[H]
\small
  \begin{center}
    \renewcommand{\arraystretch}{1.3}
    \begin{tabular}{c|cccccccc} \hline
      & $\frac{m_u}{m_t}{\times10^{6}}$ & $\frac{m_c}{m_t}{\times10^3}$ & $\frac{m_d}{m_b}{\times10^4}$ & $\frac{m_s}{m_b}{\times10^2}$ & $|V_{\textrm{CKM}}^{us}|$ & $|V_{\textrm{CKM}}^{cb}|$ & $|V_{\textrm{CKM}}^{ub}|$ & $J_{\textrm{CP}}{\times 10^5}$ \\ \hline
      obtained values & 10.8 & 1.06 & 18.2 & 3.40 & 0.228 & 0.0163 & 0.00441 & 0 \\ \hline
      GUT scale values & 5.39 & 2.80 & 9.21 & 1.82 & 0.225 & 0.0400 & 0.00353 & 2.80 \\
      $1\sigma$ errors & $\pm 1.68$ & $\pm 0.12$ & $\pm 1.02$ & $\pm 0.10$ & $\pm 0.0007$ & $\pm 0.0008$ & $\pm 0.00013$ & $^{+0.14}_{-0.12}$ \\ \hline
    \end{tabular}
  \end{center}
  \caption{The mass ratios of the quarks and the absolute values of the CKM matrix elements at the benchmark point $(\omega_1, \omega_2, \omega_3) = (1.3i, 0.6i, 0)$.
GUT scale values at $2\times 10^{16}$ GeV with $\tan \beta=5$ \cite{Antusch:2013jca,Bjorkeroth:2015ora} and $1\sigma$ errors are shown.}
\label{tab:fitting_one}
\normalsize
\end{table}
These results can realize orders of mass ratios and mixing angles.
Therefore, we can expect that ${\cal O}(1)$ deviations of parameters $\alpha^{ij}$ and $\beta^{ij}$ yield more realistic results.
Also we need to generate non-vanishing CP phase.

For CP violation, we begin to deviate $\omega_2$ from the imaginary axis\footnote{CP violation does not occur without $\textrm{Re}\Omega$ since the CP transformation is given by $\Omega\to -\Omega^*$ \cite{Baur:2019iai,Baur:2019kwi,Novichkov:2019sqv}.}.
We choose the moduli,
\begin{align}
(\omega_1, \omega_2, \omega_3) = (1.3i, 0.6i+0.86, 0),
\end{align}
where
\begin{align}
(q_1,q_2,q_3) = (0.360, 0.624e^{0.674i}, 1).
\end{align}
Next, to get more realistic mass ratios and mixing angles, we deviate $\alpha^{ij}$ and $\beta^{ij}$ by ${\cal O}(1)$,
\begin{align}
&\begin{pmatrix}
\text{-} & \text{-} & \alpha^{13} \\
\text{-} & \alpha^{22} & \alpha^{23} \\
\alpha^{31} & \alpha^{32} & \alpha^{33} \\
\end{pmatrix}
=
\begin{pmatrix}
\text{-} & \text{-} & 1.16 \\
\text{-} & 1.72 & -3.03 \\
1.00 & -1.36 & 1.00 \\
\end{pmatrix}, \quad
\begin{pmatrix}
\beta^{11} & \beta^{12} & \beta^{13} \\
\beta^{21} & \beta^{22} & \beta^{23} \\
\beta^{31} & \text{-} & \beta^{33} \\
\end{pmatrix}
=
\begin{pmatrix}
1.46 & 3.39 & 1.66 \\
3.44 & -1.00 & -2.89 \\
2.08 & \text{-} & -1.27 \\
\end{pmatrix}.
\end{align}
They lead to the quark mass ratios,
\begin{align}
&(m_u, m_c, m_t)/m_t = (8.25\times 10^{-6}, 2.70\times 10^{-3}, 1), \\
&(m_d, m_s, m_b)/m_b = (1.01\times 10^{-3}, 2.02\times 10^{-2}, 1),
\end{align}
absolute values of the CKM matrix elements,
\begin{align}
|V_{\textrm{CKM}}| =
\begin{pmatrix}
0.975 & 0.224 & 0.00340 \\
0.224 & 0.974 & 0.0392 \\
0.00809 & 0.0385 & 0.999 \\
\end{pmatrix},
\end{align}
and the value of Jarlskog invariant,
\begin{align}
J_{\textrm{CP}} = |\textrm{Im}(V_{\textrm{CKM}}^{us}V_{\textrm{CKM}}^{cb}(V_{\textrm{CKM}}^{ub}V_{\textrm{CKM}}^{cs})^*)| = 2.68\times 10^{-5}.
\end{align}
Results are summarized in Table \ref{tab:fitting}.
\begin{table}[H]
\small
  \begin{center}
    \renewcommand{\arraystretch}{1.3}
    \begin{tabular}{c|cccccccc} \hline
      & $\frac{m_u}{m_t}{\times10^{6}}$ & $\frac{m_c}{m_t}{\times10^3}$ & $\frac{m_d}{m_b}{\times10^4}$ & $\frac{m_s}{m_b}{\times10^2}$ & $|V_{\textrm{CKM}}^{us}|$ & $|V_{\textrm{CKM}}^{cb}|$ & $|V_{\textrm{CKM}}^{ub}|$ & $J_{\textrm{CP}}{\times 10^5}$ \\ \hline
      obtained values & 8.25 & 2.70 & 10.1 & 2.02 & 0.224 & 0.0392 & 0.00340 & 2.68 \\ \hline
      GUT scale values & 5.39 & 2.80 & 9.21 & 1.82 & 0.225 & 0.0400 & 0.00353 & 2.80 \\
      $1\sigma$ errors & $\pm 1.68$ & $\pm 0.12$ & $\pm 1.02$ & $\pm 0.10$ & $\pm 0.0007$ & $\pm 0.0008$ & $\pm 0.00013$ & $^{+0.14}_{-0.12}$ \\ \hline
    \end{tabular}
  \end{center}
  \caption{The mass ratios of the quarks and the absolute values of the CKM matrix elements at the benchmark point $(\omega_1, \omega_2, \omega_3) = (1.3i, 0.6i+0.86, 0)$.
GUT scale values at $2\times 10^{16}$ GeV with $\tan \beta=5$ \cite{Antusch:2013jca,Bjorkeroth:2015ora} and $1\sigma$ errors are shown.}
\label{tab:fitting}
\normalsize
\end{table}
Thus quark flavor observations can be realized without fine-tuning of coupling constants in the vicinity of invariant moduli $\Omega_{T_I}$.

Before ending this section, we comment on the CP violation in our model.
At $\omega_1\sim i\infty$, $\omega_2\sim i\infty$ and $\omega_3=0$, up and down quark mass matrices are evaluated by Eqs.~(\ref{eq:up_mass_mtx}) and (\ref{eq:down_mass_mtx}).
Then phase factors of $q_1$  in mass matrices can be canceled as follows.
We introduce the basis transformations,
\begin{align}
M_u \to U_Q^\dagger M_u U_{u_R}, \quad M_d \to U_Q^\dagger M_d U_{d_R},
\end{align}
where
\begin{align}
&U_Q =
\begin{pmatrix}
 e^{4i\textrm{Arg}(q_1)} &&\\
&  e^{4i\textrm{Arg}(q_1)} &\\
&& 1 \\
\end{pmatrix}, \\
&U_{u_R} =
\begin{pmatrix}
e^{-2i\textrm{Arg}(q_1)} &&\\
& 1 &\\
&& 1 \\
\end{pmatrix}, \\
&U_{d_R} =
\begin{pmatrix}
1 &&\\
& e^{2i\textrm{Arg}(q_1)} &\\
&& 1 \\
\end{pmatrix}.
\end{align}
Then, mass matrices are written by 
\begin{align}
&U_Q^\dagger M_u U_{u_R} \notag \\
&\simeq y_uv_u
\begin{pmatrix}
0 & 0 & C^{-1/2}\alpha^{13}48|q_1|^4q_2^{14} \\
0 & C^{-1/2}\alpha^{22}48|q_1|^4q_2^{14} & -C^{1/2}\alpha^{23}\frac{64}{\sqrt{3}}|q_1|^4q_2^{18} \\
C^{-5/4}\alpha^{31}4\sqrt{3}|q_1|^2q_2^7 & -C^{-1}\alpha^{32}\frac{4\sqrt{3}}{3}q_2^4 & \alpha^{33}\frac{1}{4\sqrt{6}} \\
\end{pmatrix}, \\
&U_Q^\dagger M_d U_{d_R} \notag \\
&\simeq y_d v_d
\begin{pmatrix}
C^{-1/2}\beta^{11}48|q_1|^4q_2^{14} & C^{-5/4}\beta^{12}4\sqrt{3}|q_1|^2q_2^7 & C^{-1/2}\beta^{13}48|q_1|^4q_2^{14} \\
-C^{1/2}\beta^{21}\frac{64}{\sqrt{3}}|q_1|^4q_2^{18} & -C^{-1/4}\beta^{22}16|q_1|^2q_2^{11} & -C^{1/2}\beta^{23}\frac{64}{\sqrt{3}}|q_1|^4q_2^{18} \\
\beta^{31}\frac{1}{4\sqrt{6}} & 0 &\beta^{33}\frac{1}{4\sqrt{6}} \\
\end{pmatrix}.
\end{align}
Consequently phase factors of $q_1$  in mass matrices are completely canceled \footnote{Similar behaviors at fixed points were studied in Refs.~\cite{Kobayashi:2019uyt,Kikuchi:2022geu}.}.
That is, $\textrm{Re}\omega_1$ cannot cause the CP violation.
On the other hand, we cannot cancel phase factors of $q_2$ by basis transformations.
Let us consider the further basis transformations,
\begin{align}
U_Q^\dagger M_u U_{u_R} \to \hat{U}_Q^\dagger U_Q^\dagger M_u U_{u_R} \hat{U}_{u_R}, \quad U_Q^\dagger M_d U_{d_R} \to \hat{U}_Q^\dagger U_Q^\dagger M_d U_{d_R} \hat{U}_{d_R},
\end{align}
where
\begin{align}
&\hat{U}_Q =
\begin{pmatrix}
e^{14i\textrm{Arg}(q_2)} &&\\
& e^{18i\textrm{Arg}(q_2)} &\\
&& 1 \\
\end{pmatrix}, \\
&\hat{U}_{u_R} =
\begin{pmatrix}
e^{-7i\textrm{Arg}(q_2)} &&\\
& e^{4i\textrm{Arg}(q_2)} &\\
&& 1 \\
\end{pmatrix}, \\
&\hat{U}_{d_R} =
\begin{pmatrix}
1 &&\\
& e^{7i\textrm{Arg}(q_2)} &\\
&& 1 \\
\end{pmatrix}.
\end{align}
Then mass matrices are written by
\begin{align}
&\hat{U}_Q^\dagger U_Q^\dagger M_u U_{u_R} \hat{U}_{u_R} \notag \\
&\simeq y_u v_u
\begin{pmatrix}
0 & 0 & C^{-1/2}\alpha^{13}48|q_1|^4|q_2|^{14} \\
0 & C^{-1/2}\alpha^{22}48|q_1|^4|q_2|^{14} & -C^{1/2}\alpha^{23}\frac{64}{\sqrt{3}}|q_1|^4|q_2|^{18} \\
C^{-5/4}\alpha^{31}4\sqrt{3}|q_1|^2|q_2|^7 & -C^{-1}\alpha^{32}\frac{4\sqrt{3}}{3}|q_2|^4 e^{8i\textrm{Ang}(q_2)} &\alpha^{33}\frac{1}{4\sqrt{6}} \\
\end{pmatrix}, \\
&\hat{U}_Q^\dagger U_Q^\dagger M_d U_{d_R} \hat{U}_{d_R} \notag \\
&\simeq y_d v_d
\begin{pmatrix}
C^{-1/2}\beta^{11}48|q_1|^4|q_2|^{14} & C^{-5/4}\beta^{12}4\sqrt{3}|q_1|^2|q_2|^7 & C^{-1/2}\beta^{13}48|q_1|^4|q_2|^{14} \\
-C^{1/2}\beta^{21}\frac{64}{\sqrt{3}}|q_1|^4|q_2|^{18} & -C^{-1/4}\beta^{22}16|q_1|^2|q_2|^{11} & -C^{1/2}\beta^{23}\frac{64}{\sqrt{3}}|q_1|^4|q_2|^{18} \\
\beta^{31}\frac{1}{4\sqrt{6}} & 0 & \beta^{33}\frac{1}{4\sqrt{6}} \\
\end{pmatrix}.
\end{align}
After the basis transformations, the phase factor $e^{8i\textrm{Arg}(q_2)}=e^{2\pi i\textrm{Re}\omega_2}$ survives in (3,2) element of $M_u$.
Therefore CP violation can be caused in the region $1/2>|\textrm{Re}\omega_2|>0$ only by the moduli, even if 
all of the coefficients $\alpha^{ij}$ and $\beta^{ij}$ are real.
Indeed, sufficient CP violation can be obtained by the deviation of $\omega_2$ from the imaginary axis as we have seen in the numerical example.
In this sense $\omega_1$ works on mass hierarchies while $\omega_2$ works on the CP violation in our model.
Thus, the Siegel modular forms with the moduli controled by multi parameters have the possibility realizing both mass hierarchies and the CP violation without the fine-tuning of coupling constants.
This is an attractive point of the Siegel modular forms for the subgroup of $Sp(6,\mathbb{Z})$.


\section{Conclusion}
\label{sec:conclusion}

We have constructed the Siegel modular forms for the subgroup of $Sp(6,\mathbb{Z})$ from zero-modes on the magnetized $T^6$.
Zero-modes on $T^6$ have $Sp(6,\mathbb{Z})$ modular symmetry as the geometrical symmetry.
They are transformed by unitary representations of the subgroup of $Sp(6,\mathbb{Z})$, $\rho(\gamma)$, in Eqs.~(\ref{eq:rho_S}) and (\ref{eq:rho_T}).
Thus zero-modes at $\vec{z}=0$ are the Siegel modular forms for the subgroup of $Sp(6,\mathbb{Z})$.
On the other hand, $T$-symmetries on zero-modes can break depending on the magnetic flux structures of $N$ matrix.
When $N$ matrix possessing non-degenerate eigenvalues, zero-modes have three $T$-symmetries although $Sp(6,\mathbb{Z})$ contains six $T$-symmetries.
Also the consistency condition of $S$-symmetry on zero-modes requires $(N\Omega)^T=N\Omega$, $N=N^T$ and $\Omega=\Omega^T$.
They restrict the structure of $\Omega$ to be commutable to $N$ matrix.

As constructing examples of the Siegel modular forms, we have studied zero-modes with $N$ matrices, which 
have three different eigenvalues.
Therefore only three $T$-symmetries denoted as $T_I$, $T_{II}$ and $T_{III}$ survive in zero-modes.
$\Omega$ which is consistent with $S$-symmetry on zero-modes is given by Eq.~(\ref{eq:omega_para}).

Under $S$, $T_I$, $T_{II}$ and $T_{III}$-transformations, zero-modes are transformed by unitary representations 
of finite subgroups of $Sp(6,\mathbb{Z})$.
We have shown several examples. 
One of examples is the group $\widetilde{\Delta}(96)\simeq \Delta(48)\rtimes Z_8$.
The unitary representations in Eq.~(\ref{eq:rhos}) form the group $\widetilde{\Delta}(96)\simeq \Delta(48)\rtimes Z_8$.
Thus zero-modes at $\vec{z}=0$ are the Siegel modular forms of weight 1/2 for $\widetilde{\Delta}(96)$.
Higher weight Siegel modular forms can be constructed by the tensor products of them.
The dimension of the Siegel modular forms of weight $k$ is $_{2k+2}C_2$.

We have studied the numerical example of the quark flavor model using the Siegel modular forms for $\widetilde{\Delta}(96)$.
To make our analysis simple we have used singlet Siegel modular forms for the model building.
To reproduce the large mass hierarchies of quarks, we concentrate on the vicinity of the cusp $\Omega_{T_I}$.
When the moduli $\Omega$ lies on the vicinity of $\Omega_{T_I}$, the Siegel modular forms take hierarchical values depending on its $T_I$-charge.
Actually, near the cusp, $(\omega_1,\omega_2,\omega_3)=(1.3i, 0.6i, 0)$, quark mass ratios as well as the absolute values of the CKM matrix elements can be realized up to ${\cal O}(1)$ although CP violation is not induced.
In order to avoid the fine-tuning by coupling constants in mass matrices, we restricted $\alpha$, $\beta=\pm 1$ in the numerical example.
The ${\cal O}(1)$ deviations of $\alpha$ and $\beta$, and the deviation of $\omega_2$ from the imaginary axis can lead to more realistic results including Jarlskog invariant as shown in Table \ref{tab:fitting}.

In the numerical example, $\omega_1$ cannot induce the CP violation.
This is because the phase transformations for the mass matrices can completely cancel the phase factor of $q_1$.
On the other hand, the phase factor of $q_2$ survives after the phase transformations.
Therefore in Table \ref{tab:fitting} we deviated $\omega_2$ to induce the CP violation.
The deviation of $\omega_1$ from the cusp ($T_I$ symmetric point) generates the large mass hierarchies and one of $\omega_2$ from the imaginary axis (CP symmetric points) generates the CP phase.
Thus, the Siegel modular forms with the moduli controlled by multi parameters give the further possibilities realizing both large mass hierarchies and the CP violation without the fine-tuning of coupling constants.

The deviations of the moduli from the (modular or CP) symmetric points are important in our model.
The moduli stabilization leading to such deviation is the key issue \footnote{See for moduli stabilization in moduli flavor models Refs.~\cite{Kobayashi:2019xvz,Ishiguro:2020tmo,Abe:2020vmv,Novichkov:2022wvg,Ishiguro:2022pde,Knapp-Perez:2023nty,King:2023snq,Kobayashi:2023spx}.
See also Ref.~\cite{Kikuchi:2023uqo}.}.
We leave it for future study.

\vspace{1.5 cm}
\noindent
{\large\bf Acknowledgement}\\

This work was supported by JSPS KAKENHI Grant Numbers JP22KJ0047 (S. K.) and JP23K03375 (T. K.), and JST SPRING Grant Number JPMJSP2119(K. N. and S. T.).


\appendix
\section*{Appendix}


\section{Group theory of $\widetilde{\Delta}(96)$}
\label{app:group_theory}

In this section, we summarize group theory of $\widetilde{\Delta}(96)\simeq \Delta(48)\rtimes Z_8$.
The order of $\widetilde{\Delta}(96)$ is 384 and the number of conjugacy class is 40.
There are two generators, $S$ and $T$, satisfying the algebraic relations,
\begin{align}
S^2=-i\mathbb{I}, ~(ST)^3 = \mathbb{I},~T^{8} = (S^{-1}T^{-1}ST)^3 = \mathbb{I}, ~S^2T = TS^2. \label{eq:algebra_S_T}
\end{align}
In Ref.~\cite{Kikuchi:2021ogn}, it was shown that these algebraic relations correspond to ones of $\widetilde{\Delta}(96)$.


\subsection{Irreducible representations}

The number of irreducible representations of $\widetilde{\Delta}(96)$ is 40.
Note that it is the same as the number of conjugacy class.
We show numbers of irreducible representations for each dimension in Table \ref{tab:num_dim_irr_reps}.
\begin{table}[H]
\centering
\begin{tabular}{c|ccccc}\hline
Dimensions & 1 & 2 & 3 & 6 & Total \\ \hline
Number of irr. reps. &  8 & 4 & 24 & 4 & 40 \\ \hline
\end{tabular}
\caption{Number of irreducible representations of $\widetilde{\Delta}(96)$.}
\label{tab:num_dim_irr_reps}
\end{table}
We show irreducible representations in Table \ref{tab:irr_reps}.

\begin{table}[H]
\centering
\begin{tabular}{c|c|c} \hline
$\bm{r}$ & $\rho(S)$ & $\rho(T)$ \\ \hline \hline
$\begin{matrix}\bm{1}_q \\(q=0,1,...,7) \\\end{matrix}$
 & $e^{-\frac{q\pi i}{4}}$
 & $e^{\frac{q\pi i}{4}}$\\ \hline

$\begin{matrix}\bm{2}_q \\(q=0,1,2,3) \\\end{matrix}$
 & $e^{-\frac{q\pi i}{4}}\begin{pmatrix}
-\frac{1}{2} & \frac{\sqrt{3}}{2} \\ 
\frac{\sqrt{3}}{2} & \frac{1}{2} \\ 
\end{pmatrix}$
 & $e^{\frac{q\pi i}{4}}\begin{pmatrix}
1 & 0 \\ 
0 & -1 \\ 
\end{pmatrix}$\\ \hline

$\begin{matrix}\bm{3}_q \\(q=0,1,...,7) \\\end{matrix}$
 & $e^{-\frac{q\pi i}{4}}\begin{pmatrix}
0 & \frac{1}{\sqrt{2}} & \frac{1}{\sqrt{2}} \\ 
\frac{1}{\sqrt{2}} & -\frac{1}{2} & \frac{1}{2} \\ 
\frac{1}{\sqrt{2}} & \frac{1}{2} & -\frac{1}{2} \\ 
\end{pmatrix}$
 & $e^{\frac{q\pi i}{4}}\begin{pmatrix}
1 & 0 & 0 \\ 
0 & 1i & 0 \\ 
0 & 0 & -1i \\ 
\end{pmatrix}$\\ \hline

$\begin{matrix}\bm{3}'_q \\(q=0,1,...,7) \\\end{matrix}$
 & $e^{-\frac{q\pi i}{4}}\begin{pmatrix}
0 & \frac{1}{\sqrt{2}}i & \frac{1}{\sqrt{2}}i \\ 
\frac{1}{\sqrt{2}}i & \frac{1}{2}i & -\frac{1}{2}i \\ 
\frac{1}{\sqrt{2}}i & -\frac{1}{2}i & \frac{1}{2}i \\ 
\end{pmatrix}$
 & $e^{\frac{q\pi i}{4}}\begin{pmatrix}
1 & 0 & 0 \\ 
0 & e^{\frac{\pi i}{4}} & 0 \\ 
0 & 0 & e^{\frac{5\pi i}{4}} \\ 
\end{pmatrix}$\\ \hline

$\begin{matrix}\bm{\hat{3}}_q \\(q=0,1,...,7) \\\end{matrix}$
 & $e^{-\frac{q\pi i}{4}}\begin{pmatrix}
0 & \frac{1}{\sqrt{2}}i & \frac{1}{\sqrt{2}}i \\ 
\frac{1}{\sqrt{2}}i & -\frac{1}{2}i & \frac{1}{2}i \\ 
\frac{1}{\sqrt{2}}i & \frac{1}{2}i & -\frac{1}{2}i \\ 
\end{pmatrix}$
 & $e^{\frac{q\pi i}{4}}\begin{pmatrix}
1 & 0 & 0 \\ 
0 & e^{\frac{3\pi i}{4}} & 0 \\ 
0 & 0 & e^{\frac{7\pi i}{4}} \\ 
\end{pmatrix}$\\ \hline

$\begin{matrix}
\bm{6}_q \\
(q=0,1,2,3) \\
\end{matrix}$
 & $e^{-\frac{q\pi i}{4}}\begin{pmatrix}
0 & \frac{1}{2} & \frac{1}{2} & 0 & \frac{1}{2} & \frac{1}{2} \\ 
\frac{1}{2} & 0 & -\frac{1}{2} & \frac{1}{2} & 0 & \frac{1}{2} \\ 
\frac{1}{2} & -\frac{1}{2} & 0 & -\frac{1}{2} & \frac{1}{2} & 0 \\ 
0 & \frac{1}{2} & -\frac{1}{2} & 0 & \frac{1}{2} & -\frac{1}{2} \\ 
\frac{1}{2} & 0 & \frac{1}{2} & \frac{1}{2} & 0 & -\frac{1}{2} \\ 
\frac{1}{2} & \frac{1}{2} & 0 & -\frac{1}{2} & -\frac{1}{2} & 0 \\ 
\end{pmatrix}$
 & $e^{\frac{q\pi i}{4}}\begin{pmatrix}
1 & 0 & 0 & 0 & 0 & 0 \\ 
0 & e^{\frac{\pi i}{4}} & 0 & 0 & 0 & 0 \\ 
0 & 0 & e^{\frac{3\pi i}{4}} & 0 & 0 & 0 \\ 
0 & 0 & 0 & -1 & 0 & 0 \\ 
0 & 0 & 0 & 0 & e^{\frac{5\pi i}{4}} & 0 \\ 
0 & 0 & 0 & 0 & 0 & e^{\frac{7\pi i}{4}} \\ 
\end{pmatrix}$\\ \hline 
\end{tabular}
\caption{Irreducible representations of $\widetilde{\Delta}(96)$.
Number of irreducible representations is 40.}
\label{tab:irr_reps}
\end{table}


\subsection{Tensor products rules}

In Table \ref{tab:tensor}, we summarize Clebsch-Gordan (CG) coefficients of tensor products.
We have used the following notations,
\begin{align}
P_2 =
\begin{pmatrix}
0 & -1 \\
1 & 0 \\
\end{pmatrix}, \quad
P_6 =
\begin{pmatrix}
0 & 0 & 0 & 1 & 0 & 0 \\
0 & 0 & 0 & 0 & -1 & 0 \\
0 & 0 & 0 & 0 & 0 & 1 \\
1 & 0 & 0 & 0 & 0 & 0 \\
0 & -1 & 0 & 0 & 0 & 0 \\
0 & 0 & 1 & 0 & 0 & 0 \\
\end{pmatrix}.
\end{align}
We also use a notation $int(x)$ to express a integral part of real number $x$.
\small
\begin{longtable}{c} 
\caption{CG coefficients of tensor products.}
\label{tab:tensor} \\\hline
$\begin{matrix}
\bm{1}_p \oplus \bm{1}_q = \bm{1}_r \\ 
\alpha^1\beta^1 \\ 
\end{matrix}$ \\ \hline
$\begin{matrix}
\bm{2}_p \otimes \bm{1}_q = \bm{2}_r \\
P_2^{int(\frac{r}{4})}
\begin{pmatrix}
\alpha^1\beta^1 \\ \alpha^2\beta^1 \\
\end{pmatrix} \\ 
\end{matrix}$ \\ \hline
$\begin{matrix}
\bm{3}_p \otimes \bm{1}_q = \bm{3}_r \\
\bm{3}'_p \otimes \bm{1}_q = \bm{3}'_r \\
\bm{\hat{3}}_p \otimes \bm{1}_q = \bm{\hat{3}}_r \\
\begin{pmatrix}
\alpha^1\beta^1 \\ \alpha^2\beta^1 \\ \alpha^3\beta^1 \\
\end{pmatrix} \\
\end{matrix}$ \\ \hline
$\begin{matrix}
\bm{6}_p \otimes \bm{1}_q = \bm{6}_r \\
P_6^{int(\frac{r}{4})}\begin{pmatrix}
\alpha^1\beta^1 \\ \alpha^2\beta^1 \\ \alpha^3\beta^1 \\ \alpha^4\beta^1 \\ \alpha^5\beta^1 \\ \alpha^6\beta^1 \\
\end{pmatrix} \\
\end{matrix}$ \\ \hline
$\begin{matrix}
\bm{2}_p\otimes \bm{2}_q = \bm{1}_r \oplus \bm{1}_{r+4} \oplus \bm{2}_r \\
\frac{1}{\sqrt{2}}(\alpha^1\beta^1+\alpha^2\beta^2) 
\oplus\frac{1}{\sqrt{2}}(\alpha^1\beta^2-\alpha^2\beta^1) 
\oplus P_2^{int(\frac{r}{4})}\frac{1}{\sqrt{2}}\begin{pmatrix}\alpha^1\beta^1-\alpha^2\beta^2 \\
-\alpha^1\beta^2-\alpha^2\beta^1 \\
\end{pmatrix} \\
\end{matrix}$ \\ \hline 
$\begin{matrix}
\bm{3}_p\otimes \bm{2}_q = \bm{3}_{r} \oplus \bm{3}_{r+4} \\ 
\bm{\hat{3}}_p\otimes \bm{2}_q = \bm{\hat{3}}_{r} \oplus \bm{\hat{3}}_{r+4} \\
\bm{3}'_p\otimes \bm{2}_q = \bm{3}'_{r} \oplus \bm{3}'_{r+4} \\
\begin{pmatrix}
\alpha^1\beta^1 \\
-\frac{1}{2}\alpha^2\beta^1+\frac{\sqrt{3}}{2}\alpha^3\beta^2 \\
\frac{\sqrt{3}}{2}\alpha^2\beta^2-\frac{1}{2}\alpha^3\beta^1 \\
\end{pmatrix}
\oplus\begin{pmatrix}
\alpha^1\beta^2 \\
-\frac{1}{2}\alpha^2\beta^2-\frac{\sqrt{3}}{2}\alpha^3\beta^1 \\
-\frac{\sqrt{3}}{2}\alpha^2\beta^1-\frac{1}{2}\alpha^3\beta^2 \\
\end{pmatrix} \\
\end{matrix}$ \\ \hline
$\begin{matrix}
\bm{6}_p\otimes \bm{2}_q = \bm{6}_{r} \oplus \bm{6}_{r} \\
P_6^{int(\frac{r}{4})}
\begin{pmatrix}
\alpha^1\beta^1 \\
-\frac{1}{2}\alpha^2\beta^1+\frac{\sqrt{3}}{2}\alpha^5\beta^2 \\
-\frac{1}{2}\alpha^3\beta^1+\frac{\sqrt{3}}{2}\alpha^6\beta^2 \\
\alpha^4\beta^1 \\
\frac{\sqrt{3}}{2}\alpha^2\beta^2-\frac{1}{2}\alpha^5\beta^1 \\
\frac{\sqrt{3}}{2}\alpha^3\beta^2-\frac{1}{2}\alpha^6\beta^1 \\
\end{pmatrix} 
\oplus P_6^{int(\frac{r}{4})}\begin{pmatrix}
\alpha^4\beta^2 \\
\frac{\sqrt{3}}{2}\alpha^2\beta^1+\frac{1}{2}\alpha^5\beta^2 \\
-\frac{\sqrt{3}}{2}\alpha^3\beta^1-\frac{1}{2}\alpha^6\beta^2 \\
\alpha^1\beta^2 \\
\frac{1}{2}\alpha^2\beta^2+\frac{\sqrt{3}}{2}\alpha^5\beta^1 \\
-\frac{1}{2}\alpha^3\beta^2-\frac{\sqrt{3}}{2}\alpha^6\beta^1 \\
\end{pmatrix}
\end{matrix}$ \\ \hline
$\begin{matrix}
\bm{3}_p \otimes \bm{3}_q = \bm{1}_r \oplus \bm{2}_r \oplus \bm{3}_r \oplus \bm{3}_{r+4} \\
\left(\begin{array}{c}\frac{\sqrt{3}}{3}\alpha^1\beta^1+\frac{\sqrt{3}}{3}\alpha^2\beta^3+\frac{\sqrt{3}}{3}\alpha^3\beta^2 \\
\end{array}\right) 
\oplus P_2^{int(\frac{r}{4})}\left(\begin{array}{c}\frac{\sqrt{6}}{3}\alpha^1\beta^1-\frac{\sqrt{6}}{6}\alpha^2\beta^3-\frac{\sqrt{6}}{6}\alpha^3\beta^2 \\
\frac{1}{\sqrt{2}}\alpha^2\beta^2+\frac{1}{\sqrt{2}}\alpha^3\beta^3 \\
\end{array}\right) \\
\oplus \left(\begin{array}{c}\frac{1}{\sqrt{2}}\alpha^2\beta^3-\frac{1}{\sqrt{2}}\alpha^3\beta^2 \\
\frac{1}{\sqrt{2}}\alpha^1\beta^2-\frac{1}{\sqrt{2}}\alpha^2\beta^1 \\
-\frac{1}{\sqrt{2}}\alpha^1\beta^3+\frac{1}{\sqrt{2}}\alpha^3\beta^1 \\
\end{array}\right) 
\oplus \left(\begin{array}{c}\frac{1}{\sqrt{2}}\alpha^2\beta^2-\frac{1}{\sqrt{2}}\alpha^3\beta^3 \\
-\frac{1}{\sqrt{2}}\alpha^1\beta^3-\frac{1}{\sqrt{2}}\alpha^3\beta^1 \\
\frac{1}{\sqrt{2}}\alpha^1\beta^2+\frac{1}{\sqrt{2}}\alpha^2\beta^1 \\
\end{array}\right) \\
\end{matrix}$ \\ \hline
$\begin{matrix}
\bm{3}'_p \otimes \bm{3}'_q = \bm{3}_r \oplus \bm{\hat{3}}_{r+2} \oplus \bm{\hat{3}}_{r+6} \\
\left(\begin{array}{c}
\alpha^1\beta^1 \\
-\frac{1}{\sqrt{2}}\alpha^2\beta^2-\frac{1}{\sqrt{2}}\alpha^3\beta^3 \\
-\frac{1}{\sqrt{2}}\alpha^2\beta^3-\frac{1}{\sqrt{2}}\alpha^3\beta^2 \\
\end{array}\right) 
\oplus\left(\begin{array}{c}
\frac{1}{\sqrt{2}}\alpha^2\beta^2-\frac{1}{\sqrt{2}}\alpha^3\beta^3 \\
\frac{1}{\sqrt{2}}\alpha^1\beta^3+\frac{1}{\sqrt{2}}\alpha^3\beta^1 \\
-\frac{1}{\sqrt{2}}\alpha^1\beta^2-\frac{1}{\sqrt{2}}\alpha^2\beta^1 \\
\end{array}\right) 
\oplus\left(\begin{array}{c}
\frac{1}{\sqrt{2}}\alpha^2\beta^3-\frac{1}{\sqrt{2}}\alpha^3\beta^2 \\
-\frac{1}{\sqrt{2}}\alpha^1\beta^2+\frac{1}{\sqrt{2}}\alpha^2\beta^1 \\
\frac{1}{\sqrt{2}}\alpha^1\beta^3-\frac{1}{\sqrt{2}}\alpha^3\beta^1 \\
\end{array}\right) \\
\end{matrix}$ \\ \hline
$\begin{matrix}
\bm{\hat{3}}_p\otimes \bm{\hat{3}}_q = \bm{3}_{r} \oplus \bm{3}'_{r+2} \oplus \bm{3}'_{r+6} \\
\begin{pmatrix}
\alpha^1\beta^1 \\
-\frac{1}{\sqrt{2}}\alpha^2\beta^3-\frac{1}{\sqrt{2}}\alpha^3\beta^2 \\
-\frac{1}{\sqrt{2}}\alpha^2\beta^2-\frac{1}{\sqrt{2}}\alpha^3\beta^3 \\
\end{pmatrix} 
\oplus\begin{pmatrix}
\frac{1}{\sqrt{2}}\alpha^2\beta^3-\frac{1}{\sqrt{2}}\alpha^3\beta^2 \\
-\frac{1}{\sqrt{2}}\alpha^1\beta^2+\frac{1}{\sqrt{2}}\alpha^2\beta^1 \\
\frac{1}{\sqrt{2}}\alpha^1\beta^3-\frac{1}{\sqrt{2}}\alpha^3\beta^1 \\
\end{pmatrix} 
\oplus\begin{pmatrix}
\frac{1}{\sqrt{2}}\alpha^2\beta^2-\frac{1}{\sqrt{2}}\alpha^3\beta^3 \\
\frac{1}{\sqrt{2}}\alpha^1\beta^3+\frac{1}{\sqrt{2}}\alpha^3\beta^1 \\
-\frac{1}{\sqrt{2}}\alpha^1\beta^2-\frac{1}{\sqrt{2}}\alpha^2\beta^1 \\
\end{pmatrix} \\
\end{matrix}$ \\ \hline
$\begin{matrix}
\bm{3}'_p\otimes \bm{\hat{3}}_q = \bm{1}_{r} \oplus \bm{2}_{r} \oplus \bm{6}_{r} \\
\left(\begin{array}{c}\frac{\sqrt{3}}{3}\alpha^1\beta^1-\frac{\sqrt{3}}{3}\alpha^2\beta^3-\frac{\sqrt{3}}{3}\alpha^3\beta^2 \\
\end{array}\right) 
\oplus P_2^{int(\frac{r}{4})}\left(\begin{array}{c}
\frac{\sqrt{6}}{3}\alpha^1\beta^1+\frac{\sqrt{6}}{6}\alpha^2\beta^3+\frac{\sqrt{6}}{6}\alpha^3\beta^2 \\
-\frac{1}{\sqrt{2}}\alpha^2\beta^2-\frac{1}{\sqrt{2}}\alpha^3\beta^3 \\
\end{array}\right) \\
\oplus P_6^{int(\frac{r}{4})}\left(\begin{array}{c}
\frac{1}{\sqrt{2}}\alpha^2\beta^3-\frac{1}{\sqrt{2}}\alpha^3\beta^2 \\
-\alpha^2\beta^1 \\
-\alpha^1\beta^2 \\
\frac{1}{\sqrt{2}}\alpha^2\beta^2-\frac{1}{\sqrt{2}}\alpha^3\beta^3 \\
\alpha^3\beta^1 \\
\alpha^1\beta^3 \\
\end{array}\right) \\
\end{matrix}$ \\ \hline
$\begin{matrix}
\bm{3}'_p \otimes \bm{3}_q = \bm{\hat{3}}_r \oplus \bm{6}_{r+2} \\
\left(\begin{array}{c}
\alpha^1\beta^1 \\
\frac{1}{\sqrt{2}}\alpha^2\beta^2+\frac{1}{\sqrt{2}}\alpha^3\beta^3 \\
\frac{1}{\sqrt{2}}\alpha^2\beta^3+\frac{1}{\sqrt{2}}\alpha^3\beta^2 \\
\end{array}\right) 
\oplus P_6^{int(\frac{r+2}{4})} \left(\begin{array}{c}
\alpha^1\beta^2 \\
\frac{1}{\sqrt{2}}\alpha^2\beta^2-\frac{1}{\sqrt{2}}\alpha^3\beta^3 \\
-\alpha^3\beta^1 \\
-\alpha^1\beta^3 \\
-\frac{1}{\sqrt{2}}\alpha^2\beta^3+\frac{1}{\sqrt{2}}\alpha^3\beta^2 \\
-\alpha^2\beta^1 \\
\end{array}\right) \\
\end{matrix}$ \\ \hline
$\begin{matrix}
\bm{3}_p \otimes \bm{\hat{3}}_q = \bm{3}'_r \oplus \bm{6}_{r+2} \\
\left(\begin{array}{c}
\alpha^1\beta^1 \\
\frac{1}{\sqrt{2}}\alpha^2\beta^3+\frac{1}{\sqrt{2}}\alpha^3\beta^2 \\
\frac{1}{\sqrt{2}}\alpha^2\beta^2+\frac{1}{\sqrt{2}}\alpha^3\beta^3 \\
\end{array}\right) 
\oplus P_6^{int(\frac{r+2}{4})} \left(\begin{array}{c}
\alpha^2\beta^1 \\
-\alpha^1\beta^2 \\
\frac{1}{\sqrt{2}}\alpha^2\beta^2-\frac{1}{\sqrt{2}}\alpha^3\beta^3 \\
\alpha^3\beta^1 \\
-\alpha^1\beta^3 \\
\frac{1}{\sqrt{2}}\alpha^2\beta^3-\frac{1}{\sqrt{2}}\alpha^3\beta^2 \\
\end{array}\right) \\
\end{matrix}$ \\ \hline
$\begin{matrix}
\bm{6}_p \otimes \bm{3}_q = \bm{3}'_{r+2} \oplus \bm{3}'_{r+6} \oplus \bm{\hat{3}}_{r+2} \oplus \bm{\hat{3}}_{r+6} \oplus \bm{6}_r \\
\left(\begin{array}{c}
\frac{1}{\sqrt{2}}\alpha^1\beta^2-\frac{1}{\sqrt{2}}\alpha^4\beta^3 \\
-\frac{1}{2}\alpha^2\beta^2+\frac{1}{\sqrt{2}}\alpha^3\beta^1+\frac{1}{2}\alpha^5\beta^3 \\
\frac{1}{\sqrt{2}}\alpha^2\beta^3-\frac{1}{2}\alpha^5\beta^2+\frac{1}{2}\alpha^6\beta^1 \\
\end{array}\right) 
\oplus \left(\begin{array}{c}
\frac{1}{\sqrt{2}}\alpha^1\beta^3-\frac{1}{\sqrt{2}}\alpha^4\beta^2 \\
\frac{1}{\sqrt{2}}\alpha^2\beta^3-\frac{1}{2}\alpha^5\beta^2-\frac{1}{2}\alpha^6\beta^1 \\
-\frac{1}{2}\alpha^2\beta^2-\frac{1}{\sqrt{2}}\alpha^3\beta^1+\frac{1}{2}\alpha^5\beta^3 \\
\end{array}\right) \\
\oplus \left(\begin{array}{c}
\frac{1}{\sqrt{2}}\alpha^1\beta^2+\frac{1}{\sqrt{2}}\alpha^4\beta^3 \\
-\frac{1}{2}\alpha^3\beta^2+\frac{1}{\sqrt{2}}\alpha^5\beta^1+\frac{1}{2}\alpha^6\beta^3 \\
\frac{1}{2}\alpha^2\beta^1+\frac{1}{2}\alpha^3\beta^3-\frac{1}{\sqrt{2}}\alpha^6\beta^2 \\
\end{array}\right) 
\oplus \left(\begin{array}{c}
\frac{1}{\sqrt{2}}\alpha^1\beta^3+\frac{1}{\sqrt{2}}\alpha^4\beta^2 \\
-\frac{1}{2}\alpha^2\beta^1+\frac{1}{2}\alpha^3\beta^3-\frac{1}{\sqrt{2}}\alpha^6\beta^2 \\
-\frac{1}{2}\alpha^3\beta^2-\frac{1}{\sqrt{2}}\alpha^5\beta^1+\frac{1}{2}\alpha^6\beta^3 \\
\end{array}\right) \\
\oplus P_6^{int(\frac{r}{4})} \left(\begin{array}{c}
\alpha^1\beta^1 \\
\frac{1}{\sqrt{2}}\alpha^3\beta^3+\frac{1}{\sqrt{2}}\alpha^6\beta^2 \\
\frac{1}{\sqrt{2}}\alpha^2\beta^2+\frac{1}{\sqrt{2}}\alpha^5\beta^3 \\
-\alpha^4\beta^1 \\
\frac{1}{\sqrt{2}}\alpha^3\beta^2+\frac{1}{\sqrt{2}}\alpha^6\beta^3 \\
\frac{1}{\sqrt{2}}\alpha^2\beta^3+\frac{1}{\sqrt{2}}\alpha^5\beta^2 \\
\end{array}\right) \\
\end{matrix}$ \\ \hline
$\begin{matrix}
\bm{6}_p \otimes \bm{3}'_q = \bm{3}_{r+2} \oplus \bm{3}_{r+6} \oplus \bm{3}'_{r} \oplus \bm{3}'_{r+4} \oplus \bm{6}_{r+2} \\
\left(\begin{array}{c}
\frac{1}{\sqrt{2}}\alpha^2\beta^2+\frac{1}{\sqrt{2}}\alpha^5\beta^3 \\
\frac{1}{2}\alpha^3\beta^2-\frac{1}{\sqrt{2}}\alpha^4\beta^1+\frac{1}{2}\alpha^6\beta^3 \\
-\frac{1}{2}\alpha^1\beta^1-\frac{1}{2}\alpha^3\beta^3-\frac{1}{\sqrt{2}}\alpha^6\beta^2 \\
\end{array}\right) 
\oplus \left(\begin{array}{c}
\frac{1}{\sqrt{2}}\alpha^2\beta^3+\frac{1}{\sqrt{2}}\alpha^5\beta^2 \\
\frac{1}{2}\alpha^1\beta^1-\frac{1}{2}\alpha^3\beta^3-\frac{1}{\sqrt{2}}\alpha^6\beta^2 \\
\frac{1}{2}\alpha^3\beta^2+\frac{1}{\sqrt{2}}\alpha^4\beta^1+\frac{1}{2}\alpha^6\beta^3 \\
\end{array}\right) \\
\oplus \left(\begin{array}{c}
\frac{1}{\sqrt{2}}\alpha^3\beta^3-\frac{1}{\sqrt{2}}\alpha^6\beta^2 \\
-\frac{1}{2}\alpha^1\beta^2-\frac{1}{\sqrt{2}}\alpha^2\beta^1-\frac{1}{2}\alpha^4\beta^3 \\
\frac{1}{\sqrt{2}}\alpha^1\beta^3+\frac{1}{2}\alpha^4\beta^2+\frac{1}{2}\alpha^5\beta^1 \\
\end{array}\right) 
\oplus \left(\begin{array}{c}
\frac{1}{\sqrt{2}}\alpha^3\beta^2-\frac{1}{\sqrt{2}}\alpha^6\beta^3 \\
\frac{1}{2}\alpha^1\beta^3+\frac{1}{2}\alpha^4\beta^2-\frac{1}{\sqrt{2}}\alpha^5\beta^1 \\
-\frac{1}{2}\alpha^1\beta^2+\frac{1}{\sqrt{2}}\alpha^2\beta^1-\frac{1}{2}\alpha^4\beta^3 \\
\end{array}\right) \\
\oplus P_6^{int(\frac{r+2}{4})}\left(\begin{array}{c}
\frac{1}{\sqrt{2}}\alpha^2\beta^2-\frac{1}{\sqrt{2}}\alpha^5\beta^3 \\
\alpha^3\beta^1 \\
\frac{1}{\sqrt{2}}\alpha^1\beta^3-\frac{1}{\sqrt{2}}\alpha^4\beta^2 \\
\frac{1}{\sqrt{2}}\alpha^2\beta^3-\frac{1}{\sqrt{2}}\alpha^5\beta^2 \\
-\alpha^6\beta^1 \\
-\frac{1}{\sqrt{2}}\alpha^1\beta^2+\frac{1}{\sqrt{2}}\alpha^4\beta^3 \\
\end{array}\right) \\
\end{matrix}$ \\ \hline
$\begin{matrix}
\bm{6}_p \otimes \bm{\hat{3}}_q = \bm{3}_{r+2} \oplus \bm{3}_{r+6} \oplus \bm{\hat{3}}_{r} \oplus \bm{\hat{3}}_{r+4} \oplus \bm{6}_{r+2} \\
\left(\begin{array}{c}\frac{1}{\sqrt{2}}\alpha^3\beta^3+\frac{1}{\sqrt{2}}\alpha^6\beta^2 \\
\frac{1}{2}\alpha^2\beta^2+\frac{1}{\sqrt{2}}\alpha^4\beta^1+\frac{1}{2}\alpha^5\beta^3 \\
-\frac{1}{\sqrt{2}}\alpha^1\beta^1-\frac{1}{2}\alpha^2\beta^3-\frac{1}{2}\alpha^5\beta^2 \\
\end{array}\right) 
\oplus\left(\begin{array}{c}\frac{1}{\sqrt{2}}\alpha^3\beta^2+\frac{1}{\sqrt{2}}\alpha^6\beta^3 \\
\frac{1}{\sqrt{2}}\alpha^1\beta^1-\frac{1}{2}\alpha^2\beta^3-\frac{1}{2}\alpha^5\beta^2 \\
\frac{1}{2}\alpha^2\beta^2-\frac{1}{\sqrt{2}}\alpha^4\beta^1+\frac{1}{2}\alpha^5\beta^3 \\
\end{array}\right) \\
\oplus\left(\begin{array}{c}\frac{1}{\sqrt{2}}\alpha^2\beta^3-\frac{1}{\sqrt{2}}\alpha^5\beta^2 \\
\frac{1}{2}\alpha^1\beta^2-\frac{1}{\sqrt{2}}\alpha^3\beta^1-\frac{1}{2}\alpha^4\beta^3 \\
-\frac{1}{2}\alpha^1\beta^3+\frac{1}{2}\alpha^4\beta^2+\frac{1}{\sqrt{2}}\alpha^6\beta^1 \\
\end{array}\right) 
\oplus\left(\begin{array}{c}\frac{1}{\sqrt{2}}\alpha^2\beta^2-\frac{1}{\sqrt{2}}\alpha^5\beta^3 \\
-\frac{1}{2}\alpha^1\beta^3+\frac{1}{2}\alpha^4\beta^2-\frac{1}{\sqrt{2}}\alpha^6\beta^1 \\
\frac{1}{2}\alpha^1\beta^2+\frac{1}{\sqrt{2}}\alpha^3\beta^1-\frac{1}{2}\alpha^4\beta^3 \\
\end{array}\right) \\
\oplus P_6^{int(\frac{r+2}{4})}\left(\begin{array}{c}\frac{1}{\sqrt{2}}\alpha^3\beta^3-\frac{1}{\sqrt{2}}\alpha^6\beta^2 \\
-\frac{1}{\sqrt{2}}\alpha^1\beta^2-\frac{1}{\sqrt{2}}\alpha^4\beta^3 \\
-\alpha^5\beta^1 \\
-\frac{1}{\sqrt{2}}\alpha^3\beta^2+\frac{1}{\sqrt{2}}\alpha^6\beta^3 \\
\frac{1}{\sqrt{2}}\alpha^1\beta^3+\frac{1}{\sqrt{2}}\alpha^4\beta^2 \\
\alpha^2\beta^1 \\
\end{array}\right) \\
\end{matrix}$ \\ \hline
$\begin{matrix}
\bm{6}_p \otimes \bm{6}_q = \bm{1}_{r} \oplus \bm{1}_{r+4} \oplus \bm{2}_{r} \oplus \bm{2}_{r} \oplus \bm{3}_{r} \oplus \bm{3}_{r+4} \oplus \bm{3}'_{r+2} \oplus \bm{3}'_{r+6} \oplus \bm{\hat{3}}_{r+2} \oplus \bm{\hat{3}}_{r+6} \oplus \bm{6}_{r} \oplus \bm{6}_{r} \\
\left(\begin{array}{c}\frac{\sqrt{6}}{6}\alpha^1\beta^1+\frac{\sqrt{6}}{6}\alpha^2\beta^6+\frac{\sqrt{6}}{6}\alpha^3\beta^5-\frac{\sqrt{6}}{6}\alpha^4\beta^4+\frac{\sqrt{6}}{6}\alpha^5\beta^3+\frac{\sqrt{6}}{6}\alpha^6\beta^2 \\
\end{array}\right) \\
\oplus \left(\begin{array}{c}\frac{\sqrt{6}}{6}\alpha^1\beta^4+\frac{\sqrt{6}}{6}\alpha^2\beta^3-\frac{\sqrt{6}}{6}\alpha^3\beta^2-\frac{\sqrt{6}}{6}\alpha^4\beta^1+\frac{\sqrt{6}}{6}\alpha^5\beta^6-\frac{\sqrt{6}}{6}\alpha^6\beta^5 \\
\end{array}\right) \\
\oplus P_2^{int(\frac{r}{4})} \left(\begin{array}{c}\frac{1}{2}\alpha^2\beta^6-\frac{1}{2}\alpha^3\beta^5+\frac{1}{2}\alpha^5\beta^3-\frac{1}{2}\alpha^6\beta^2 \\
-\frac{\sqrt{3}}{3}\alpha^1\beta^4+\frac{\sqrt{3}}{6}\alpha^2\beta^3-\frac{\sqrt{3}}{6}\alpha^3\beta^2+\frac{\sqrt{3}}{3}\alpha^4\beta^1+\frac{\sqrt{3}}{6}\alpha^5\beta^6-\frac{\sqrt{3}}{6}\alpha^6\beta^5 \\
\end{array}\right) \\
\oplus P_2^{int(\frac{r}{4})} \left(\begin{array}{c}\frac{\sqrt{3}}{3}\alpha^1\beta^1-\frac{\sqrt{3}}{6}\alpha^2\beta^6-\frac{\sqrt{3}}{6}\alpha^3\beta^5-\frac{\sqrt{3}}{3}\alpha^4\beta^4-\frac{\sqrt{3}}{6}\alpha^5\beta^3-\frac{\sqrt{3}}{6}\alpha^6\beta^2 \\
\frac{1}{2}\alpha^2\beta^3+\frac{1}{2}\alpha^3\beta^2+\frac{1}{2}\alpha^5\beta^6+\frac{1}{2}\alpha^6\beta^5 \\
\end{array}\right) \\
\oplus \left(\begin{array}{c}\frac{1}{\sqrt{2}}\alpha^1\beta^1+\frac{1}{\sqrt{2}}\alpha^4\beta^4 \\
\frac{1}{2}\alpha^2\beta^2+\frac{1}{2}\alpha^3\beta^6+\frac{1}{2}\alpha^5\beta^5+\frac{1}{2}\alpha^6\beta^3 \\
\frac{1}{2}\alpha^2\beta^5+\frac{1}{2}\alpha^3\beta^3+\frac{1}{2}\alpha^5\beta^2+\frac{1}{2}\alpha^6\beta^6 \\
\end{array}\right) 
\oplus \left(\begin{array}{c}\frac{1}{\sqrt{2}}\alpha^1\beta^4+\frac{1}{\sqrt{2}}\alpha^4\beta^1 \\
-\frac{1}{2}\alpha^2\beta^5+\frac{1}{2}\alpha^3\beta^3-\frac{1}{2}\alpha^5\beta^2+\frac{1}{2}\alpha^6\beta^6 \\
-\frac{1}{2}\alpha^2\beta^2+\frac{1}{2}\alpha^3\beta^6-\frac{1}{2}\alpha^5\beta^5+\frac{1}{2}\alpha^6\beta^3 \\
\end{array}\right) \\
\oplus \left(\begin{array}{c}\frac{1}{\sqrt{2}}\alpha^2\beta^2-\frac{1}{\sqrt{2}}\alpha^5\beta^5 \\
-\frac{1}{2}\alpha^1\beta^3-\frac{1}{2}\alpha^3\beta^1+\frac{1}{2}\alpha^4\beta^6+\frac{1}{2}\alpha^6\beta^4 \\
\frac{1}{2}\alpha^1\beta^6-\frac{1}{2}\alpha^3\beta^4-\frac{1}{2}\alpha^4\beta^3+\frac{1}{2}\alpha^6\beta^1 \\
\end{array}\right) 
\oplus \left(\begin{array}{c}\frac{1}{\sqrt{2}}\alpha^2\beta^5-\frac{1}{\sqrt{2}}\alpha^5\beta^2 \\
\frac{1}{2}\alpha^1\beta^6+\frac{1}{2}\alpha^3\beta^4-\frac{1}{2}\alpha^4\beta^3-\frac{1}{2}\alpha^6\beta^1 \\
-\frac{1}{2}\alpha^1\beta^3+\frac{1}{2}\alpha^3\beta^1+\frac{1}{2}\alpha^4\beta^6-\frac{1}{2}\alpha^6\beta^4 \\
\end{array}\right) \\
\oplus \left(\begin{array}{c}\frac{1}{\sqrt{2}}\alpha^3\beta^6-\frac{1}{\sqrt{2}}\alpha^6\beta^3 \\
-\frac{1}{2}\alpha^1\beta^5+\frac{1}{2}\alpha^2\beta^4-\frac{1}{2}\alpha^4\beta^2+\frac{1}{2}\alpha^5\beta^1 \\
\frac{1}{2}\alpha^1\beta^2-\frac{1}{2}\alpha^2\beta^1+\frac{1}{2}\alpha^4\beta^5-\frac{1}{2}\alpha^5\beta^4 \\
\end{array}\right) 
\oplus\left(\begin{array}{c}\frac{1}{\sqrt{2}}\alpha^3\beta^3-\frac{1}{\sqrt{2}}\alpha^6\beta^6 \\
\frac{1}{2}\alpha^1\beta^2+\frac{1}{2}\alpha^2\beta^1+\frac{1}{2}\alpha^4\beta^5+\frac{1}{2}\alpha^5\beta^4 \\
-\frac{1}{2}\alpha^1\beta^5-\frac{1}{2}\alpha^2\beta^4-\frac{1}{2}\alpha^4\beta^2-\frac{1}{2}\alpha^5\beta^1 \\
\end{array}\right) \\
\oplus P_6^{int(\frac{r}{4})}\left(\begin{array}{c}\frac{1}{\sqrt{2}}\alpha^3\beta^5-\frac{1}{\sqrt{2}}\alpha^6\beta^2 \\
-\frac{1}{\sqrt{2}}\alpha^2\beta^1+\frac{1}{\sqrt{2}}\alpha^5\beta^4 \\
\frac{1}{\sqrt{2}}\alpha^1\beta^3+\frac{1}{\sqrt{2}}\alpha^4\beta^6 \\
\frac{1}{\sqrt{2}}\alpha^3\beta^2-\frac{1}{\sqrt{2}}\alpha^6\beta^5 \\
-\frac{1}{\sqrt{2}}\alpha^2\beta^4+\frac{1}{\sqrt{2}}\alpha^5\beta^1 \\
-\frac{1}{\sqrt{2}}\alpha^1\beta^6-\frac{1}{\sqrt{2}}\alpha^4\beta^3 \\
\end{array}\right) 
\oplus P_6^{int(\frac{r}{4})}\left(\begin{array}{c}\frac{1}{\sqrt{2}}\alpha^2\beta^6-\frac{1}{\sqrt{2}}\alpha^5\beta^3 \\
\frac{1}{\sqrt{2}}\alpha^1\beta^2-\frac{1}{\sqrt{2}}\alpha^4\beta^5 \\
-\frac{1}{\sqrt{2}}\alpha^3\beta^1-\frac{1}{\sqrt{2}}\alpha^6\beta^4 \\
-\frac{1}{\sqrt{2}}\alpha^2\beta^3+\frac{1}{\sqrt{2}}\alpha^5\beta^6 \\
-\frac{1}{\sqrt{2}}\alpha^1\beta^5+\frac{1}{\sqrt{2}}\alpha^4\beta^2 \\
\frac{1}{\sqrt{2}}\alpha^3\beta^4+\frac{1}{\sqrt{2}}\alpha^6\beta^1 \\
\end{array}\right) \\
\end{matrix}$ \\ \hline
\end{longtable}
\normalsize


\section{Siegel modular forms up to weight 5}
\label{app:modularforms}

In this section, we show the Siegel modular forms of $\widetilde{\Delta}(96)$ up to weight 5.
The Siegel modular forms of half integral weights, odd integral weights and even integral weights are transformed by the groups $\widetilde{\Delta}(96)$, $\Delta'(96)$ and $\Delta(96)$, respectively.
We summarize the groups informations in Table \ref{tab:weights and groups}.
\begin{table}[H]
\centering
\renewcommand{\arraystretch}{1.3}
\begin{tabular}{ccccc} \hline
Weights & Group & Order & \# of conjugacy class & \# of singlets \\ \hline
Half integer & $\widetilde{\Delta}(96)\simeq \Delta(48)\rtimes Z_8$ & 384 & 40 & 8 \\
Odd integer & $\Delta'(96)\simeq \Delta(48)\rtimes Z_4$ & 192 & 20 & 4 \\
Even integer & $\Delta(96)\simeq \Delta(48)\rtimes Z_2$ & 96 & 10 & 2 \\ \hline
\end{tabular}
\caption{Finite groups Siegel modular forms of each weight obey.}
\label{tab:weights and groups}
\end{table}

\begin{table}[H]
\centering
\renewcommand{\arraystretch}{1.3}
\begin{tabular}{c|c|c} \hline
Weights & Dimensions & Irreducible representations of the Siegel modular forms \\ \hline
$1/2$ & 3 & $\bm{3}'_7$ \\ \hline
$1$ & 6 & $\bm{3}_6$, $\bm{\hat{3}}_0$ \\ \hline
$3/2$ & 10 & $\bm{1}_7$, $\bm{\hat{3}}_5$, $\bm{6}_3$ \\ \hline
$2$ & 15 & $\bm{1}_4$, $\bm{2}_0$, $\bm{3}_0$, $\bm{3}'_6$, $\bm{6}_0$ \\ \hline
$5/2$ & 21 & $\bm{3}_5$, $\bm{3}'_{3a}$, $\bm{3}'_{3b}$, $\bm{3}'_7$, $\bm{\hat{3}}_7$, $\bm{6}_1$ \\ \hline
$3$ & 28 & $\bm{1}_6$, $\bm{3}_{2a}$, $\bm{3}_{2b}$, $\bm{3}_6$, $\bm{3}'_0$, $\bm{\hat{3}}_0$, $\bm{\hat{3}}_{4a}$, $\bm{\hat{3}}_{4b}$, $\bm{6}_2$ \\ \hline
$7/2$ & 36 & $\bm{1}_3$, $\bm{2}_3$, $\bm{3}_7$, $\bm{3}'_5$, $\bm{\hat{3}}_{1a}$, $\bm{\hat{3}}_{1b}$, $\bm{\hat{3}}_5$, $\bm{6}_{3a}$, $\bm{6}_{3b}$, $\bm{6}_{3c}$ \\ \hline
$4$ & 45 & $\bm{1}_{0a}$, $\bm{1}_{0b}$, $\bm{2}_{0a}$, $\bm{2}_{0b}$, $\bm{3}_0$, $\bm{3}_{4a}$, $\bm{3}_{4b}$, $\bm{3}'_{2a}$, $\bm{3}'_{2b}$, $\bm{3}'_6$, $\bm{\hat{3}}_6$, $\bm{6}_{0a}$, $\bm{6}_{0b}$, $\bm{6}_{0c}$ \\ \hline
$9/2$ & 55 & $\bm{1}_5$, $\bm{3}_{1a}$, $\bm{3}_{1b}$, $\bm{3}_5$, $\bm{3}'_{3a}$, $\bm{3}'_{3b}$, $\bm{3}'_{7a}$, $\bm{3}'_{7b}$, $\bm{3}'_{7c}$, $\bm{3}'_{7d}$, $\bm{\hat{3}}_{3a}$, $\bm{\hat{3}}_{3b}$, $\bm{\hat{3}}_7$, $\bm{6}_{1a}$, $\bm{6}_{1b}$, $\bm{6}_{1c}$ \\ \hline
$5$ & 66 & 
$\begin{matrix}
\bm{1}_2, \bm{2}_2, \bm{3}_{2a}, \bm{3}_{2b}, \bm{3}_{6a}, \bm{3}_{6b}, \bm{3}_{6c}, \bm{3}_{6d}, \bm{3}'_0,  \bm{3}'_{4a}, \bm{3}'_{4b}, \bm{\hat{3}}_{0a}, \bm{\hat{3}}_{0b}, \bm{\hat{3}}_{0c}, \bm{\hat{3}}_{0d}, \bm{\hat{3}}_{4a}, \bm{\hat{3}}_{4b}, \\ \bm{6}_{2a}, \bm{6}_{2b}, \bm{6}_{2c} 
\end{matrix}$
\\ \hline
\end{tabular}
\caption{The Siegel modular forms of integral weights for $\widetilde{\Delta}(96)$.
The dimensions of the Siegel modular forms of weight $k$ are $\frac{1}{2}(2k+2)(2k+1)$.}
\end{table}
As discussed in section \ref{sec:mf}, we find the Siegel modular forms of weight 1/2,
\begin{align*}
&Y^{(1/2)}_{\bm{3}'_7} = 
%
.
\end{align*}
Higher Siegel modular forms can be constructed by tensor products of $(\zeta,\eta,\theta)$.
From the tensor product of the Siegel modular forms of weight 1/2, we find the Siegel modular forms of weight 1,
\begin{longtable}{l}
$Y^{(1)}_{\bm{3}_6} =
\begin{pmatrix}
Y^{(1/2)}_{\bm{3}'_7}Y^{(1/2)}_{\bm{3}'_7}
\end{pmatrix}_{\bm{3}_6} =
\begin{pmatrix}
\zeta^{2} \\
-\frac{1}{\sqrt{2}}\theta^{2}-\frac{1}{\sqrt{2}}\eta^{2} \\
-\sqrt{2}\eta\theta \\
\end{pmatrix}$, \\
$Y^{(1)}_{\bm{\hat{3}}_0} =
\begin{pmatrix}
Y^{(1/2)}_{\bm{3}'_7}Y^{(1/2)}_{\bm{3}'_7}
\end{pmatrix}_{\bm{\hat{3}}_0} =
\begin{pmatrix}
-\frac{1}{\sqrt{2}}\theta^{2}+\frac{1}{\sqrt{2}}\eta^{2} \\
\sqrt{2}\zeta\theta \\
-\sqrt{2}\zeta\eta \\
\end{pmatrix}$.
\end{longtable}
From the tensor product of the Siegel modular forms of weight 1/2 and 1, we find the Siegel modular forms of weight 3/2,
\begin{longtable}{l}
$Y^{(3/2)}_{\bm{1}_7} =
\begin{pmatrix}
Y^{(1)}_{\bm{\hat{3}}_0}Y^{(1/2)}_{\bm{3}'_7}
\end{pmatrix}_{\bm{1}_7} =
\begin{pmatrix}
-\frac{\sqrt{6}}{2}\zeta\theta^{2}+\frac{\sqrt{6}}{2}\zeta\eta^{2} \\
\end{pmatrix}$, \\
$Y^{(3/2)}_{\bm{\hat{3}}_5} =
\begin{pmatrix}
Y^{(1)}_{\bm{3}_6}Y^{(1/2)}_{\bm{3}'_7}
\end{pmatrix}_{\bm{\hat{3}}_5} =
\begin{pmatrix}
\zeta^{3} \\
-\frac{3}{2}\eta\theta^{2}-\frac{1}{2}\eta^{3} \\
-\frac{1}{2}\theta^{3}-\frac{3}{2}\eta^{2}\theta \\
\end{pmatrix}$, \\
$Y^{(3/2)}_{\bm{6}_3} =
\begin{pmatrix}
Y^{(1)}_{\bm{3}_6}Y^{(1/2)}_{\bm{3}'_7}
\end{pmatrix}_{\bm{6}_3} =
\begin{pmatrix}
-\sqrt{2}\zeta\eta\theta \\
-\frac{1}{2}\theta^{3}+\frac{1}{2}\eta^{2}\theta \\
\zeta^{2}\eta \\
\frac{1}{\sqrt{2}}\zeta\theta^{2}+\frac{1}{\sqrt{2}}\zeta\eta^{2} \\
\frac{1}{2}\eta\theta^{2}-\frac{1}{2}\eta^{3} \\
\zeta^{2}\theta \\
\end{pmatrix}$.
\end{longtable}
From the tensor product of the Siegel modular forms of weight 1, we find the Siegel modular forms of weight 2,
\begin{longtable}{l}
$Y^{(2)}_{\bm{1}_4} =
\begin{pmatrix}
Y^{(1)}_{\bm{3}_6}Y^{(1)}_{\bm{3}_6}
\end{pmatrix}_{\bm{1}_4} =
\begin{pmatrix}
\frac{2\sqrt{3}}{3}\eta\theta^{3}+\frac{2\sqrt{3}}{3}\eta^{3}\theta+\frac{1}{\sqrt{3}}\zeta^{4} \\
\end{pmatrix}$, \\
$Y^{(2)}_{\bm{2}_0} =
\begin{pmatrix}
Y^{(1)}_{\bm{3}_6}Y^{(1)}_{\bm{3}_6}
\end{pmatrix}_{\bm{2}_0} =
\begin{pmatrix}
\frac{1}{2\sqrt{2}}\theta^{4}+\frac{3\sqrt{2}}{2}\eta^{2}\theta^{2}+\frac{1}{2\sqrt{2}}\eta^{4} \\
\frac{2}{\sqrt{6}}\eta\theta^{3}+\frac{2}{\sqrt{6}}\eta^{3}\theta-\frac{2}{\sqrt{6}}\zeta^{4} \\
\end{pmatrix}$, \\
$Y^{(2)}_{\bm{3}_0} =
\begin{pmatrix}
Y^{(1)}_{\bm{3}_6}Y^{(1)}_{\bm{3}_6}
\end{pmatrix}_{\bm{3}_0} =
\begin{pmatrix}
\frac{1}{2\sqrt{2}}\theta^{4}-\frac{1}{\sqrt{2}}\eta^{2}\theta^{2}+\frac{1}{2\sqrt{2}}\eta^{4} \\
2\zeta^{2}\eta\theta \\
-\zeta^{2}\theta^{2}-\zeta^{2}\eta^{2} \\
\end{pmatrix}$, \\
$Y^{(2)}_{\bm{3}'_6} =
\begin{pmatrix}
Y^{(1)}_{\bm{3}_6}Y^{(1)}_{\bm{\hat{3}}_0}
\end{pmatrix}_{\bm{3}'_6} =
\begin{pmatrix}
-\frac{1}{\sqrt{2}}\zeta^{2}\theta^{2}+\frac{1}{\sqrt{2}}\zeta^{2}\eta^{2} \\
-\frac{1}{\sqrt{2}}\zeta\eta\theta^{2}+\frac{1}{\sqrt{2}}\zeta\eta^{3} \\
-\frac{1}{\sqrt{2}}\zeta\theta^{3}+\frac{1}{\sqrt{2}}\zeta\eta^{2}\theta \\
\end{pmatrix}$, \\
$Y^{(2)}_{\bm{6}_0} =
\begin{pmatrix}
Y^{(1)}_{\bm{3}_6}Y^{(1)}_{\bm{\hat{3}}_0}
\end{pmatrix}_{\bm{6}_0} =
\begin{pmatrix}
\frac{1}{2}\theta^{4}-\frac{1}{2}\eta^{4} \\
-\sqrt{2}\zeta^{3}\theta \\
-\frac{1}{\sqrt{2}}\zeta\theta^{3}-\frac{3\sqrt{2}}{2}\zeta\eta^{2}\theta \\
\eta\theta^{3}-\eta^{3}\theta \\
\sqrt{2}\zeta^{3}\eta \\
\frac{3\sqrt{2}}{2}\zeta\eta\theta^{2}+\frac{1}{\sqrt{2}}\zeta\eta^{3} \\
\end{pmatrix}$.
\end{longtable}
From the tensor product of the Siegel modular forms of weight 1 and 3/2, we find the Siegel modular forms of weight 5/2,
\begin{longtable}{l}
$Y^{(5/2)}_{\bm{3}_5} =
\begin{pmatrix}
Y^{(3/2)}_{\bm{\hat{3}}_5}Y^{(1)}_{\bm{\hat{3}}_0}
\end{pmatrix}_{\bm{3}_5} =
\begin{pmatrix}
-\frac{1}{\sqrt{2}}\zeta^{3}\theta^{2}+\frac{1}{\sqrt{2}}\zeta^{3}\eta^{2} \\
\frac{1}{2}\zeta\theta^{4}-\frac{1}{2}\zeta\eta^{4} \\
\zeta\eta\theta^{3}-\zeta\eta^{3}\theta \\
\end{pmatrix}$, \\
$Y^{(5/2)}_{\bm{3}'_{3a}} =
\begin{pmatrix}
Y^{(3/2)}_{\bm{\hat{3}}_5}Y^{(1)}_{\bm{\hat{3}}_0}
\end{pmatrix}_{\bm{3}'_3} =
\begin{pmatrix}
-2\zeta\eta\theta^{3}-2\zeta\eta^{3}\theta \\
\frac{1}{4}\theta^{5}+\frac{1}{2}\eta^{2}\theta^{3}-\frac{3}{4}\eta^{4}\theta-\zeta^{4}\eta \\
-\frac{3}{4}\eta\theta^{4}+\frac{1}{2}\eta^{3}\theta^{2}+\frac{1}{4}\eta^{5}-\zeta^{4}\theta \\
\end{pmatrix}$, \\
$Y^{(5/2)}_{\bm{3}'_{3b}} =
\begin{pmatrix}
Y^{(3/2)}_{\bm{\hat{3}}_5}Y^{(1)}_{\bm{3}_6}
\end{pmatrix}_{\bm{3}'_3} =
\begin{pmatrix}
\zeta^{5} \\
\frac{1}{4}\theta^{5}+\frac{5}{2}\eta^{2}\theta^{3}+\frac{5}{4}\eta^{4}\theta \\
\frac{5}{4}\eta\theta^{4}+\frac{5}{2}\eta^{3}\theta^{2}+\frac{1}{4}\eta^{5} \\
\end{pmatrix}$, \\
$Y^{(5/2)}_{\bm{3}'_7} =
\begin{pmatrix}
Y^{(3/2)}_{\bm{\hat{3}}_5}Y^{(1)}_{\bm{\hat{3}}_0}
\end{pmatrix}_{\bm{3}'_7} =
\begin{pmatrix}
\frac{1}{2}\zeta\theta^{4}+3\zeta\eta^{2}\theta^{2}+\frac{1}{2}\zeta\eta^{4} \\
\frac{3}{4}\eta\theta^{4}-\frac{1}{2}\eta^{3}\theta^{2}-\frac{1}{4}\eta^{5}-\zeta^{4}\theta \\
-\frac{1}{4}\theta^{5}-\frac{1}{2}\eta^{2}\theta^{3}+\frac{3}{4}\eta^{4}\theta-\zeta^{4}\eta \\
\end{pmatrix}$, \\
$Y^{(5/2)}_{\bm{\hat{3}}_7} =
\begin{pmatrix}
Y^{(3/2)}_{\bm{6}_3}Y^{(1)}_{\bm{\hat{3}}_0}
\end{pmatrix}_{\bm{\hat{3}}_7} =
\begin{pmatrix}
-\frac{1}{2}\zeta\theta^{4}+\zeta\eta^{2}\theta^{2}-\frac{1}{2}\zeta\eta^{4} \\
\zeta^{2}\theta^{3}-\zeta^{2}\eta^{2}\theta \\
-\zeta^{2}\eta\theta^{2}+\zeta^{2}\eta^{3} \\
\end{pmatrix}$, \\
\end{longtable}
From the tensor product of the Siegel modular forms of weight 3/2, we find the Siegel modular forms of weight 3,
\begin{longtable}{l}
$Y^{(3)}_{\bm{1}_6} =
\begin{pmatrix}
Y^{(3/2)}_{\bm{1}_7}Y^{(3/2)}_{\bm{1}_7}
\end{pmatrix}_{\bm{1}_6} =
\begin{pmatrix}
\frac{3}{2}\zeta^{2}\theta^{4}-3\zeta^{2}\eta^{2}\theta^{2}+\frac{3}{2}\zeta^{2}\eta^{4} \\
\end{pmatrix}$, \\
$Y^{(3)}_{\bm{3}_{2a}} =
\begin{pmatrix}
Y^{(3/2)}_{\bm{\hat{3}}_5}Y^{(3/2)}_{\bm{\hat{3}}_5}
\end{pmatrix}_{\bm{3}_2} =
\begin{pmatrix}
\zeta^{6} \\
-\frac{3\sqrt{2}}{4}\eta\theta^{5}-\frac{5\sqrt{2}}{2}\eta^{3}\theta^{3}-\frac{3\sqrt{2}}{4}\eta^{5}\theta \\
-\frac{1}{4\sqrt{2}}\theta^{6}-\frac{15\sqrt{2}}{8}\eta^{2}\theta^{4}-\frac{15\sqrt{2}}{8}\eta^{4}\theta^{2}-\frac{1}{4\sqrt{2}}\eta^{6} \\
\end{pmatrix}$, \\
$Y^{(3)}_{\bm{3}_{2b}} =
\begin{pmatrix}
Y^{(3/2)}_{\bm{6}_3}Y^{(3/2)}_{\bm{6}_3}
\end{pmatrix}_{\bm{3}_2} =
\begin{pmatrix}
-\sqrt{2}\zeta^{2}\eta\theta^{3}-\sqrt{2}\zeta^{2}\eta^{3}\theta \\
\frac{1}{4}\eta\theta^{5}-\frac{1}{2}\eta^{3}\theta^{3}+\frac{1}{4}\eta^{5}\theta+\frac{1}{2}\zeta^{4}\theta^{2}+\frac{1}{2}\zeta^{4}\eta^{2} \\
-\frac{1}{8}\theta^{6}+\frac{1}{8}\eta^{2}\theta^{4}+\frac{1}{8}\eta^{4}\theta^{2}-\frac{1}{8}\eta^{6}+\zeta^{4}\eta\theta \\
\end{pmatrix}$, \\
$Y^{(3)}_{\bm{3}_6} =
\begin{pmatrix}
Y^{(3/2)}_{\bm{6}_3}Y^{(3/2)}_{\bm{6}_3}
\end{pmatrix}_{\bm{3}_6} =
\begin{pmatrix}
\frac{1}{2\sqrt{2}}\zeta^{2}\theta^{4}+\frac{3\sqrt{2}}{2}\zeta^{2}\eta^{2}\theta^{2}+\frac{1}{2\sqrt{2}}\zeta^{2}\eta^{4} \\
\frac{1}{8}\theta^{6}-\frac{1}{8}\eta^{2}\theta^{4}-\frac{1}{8}\eta^{4}\theta^{2}+\frac{1}{8}\eta^{6}+\zeta^{4}\eta\theta \\
-\frac{1}{4}\eta\theta^{5}+\frac{1}{2}\eta^{3}\theta^{3}-\frac{1}{4}\eta^{5}\theta+\frac{1}{2}\zeta^{4}\theta^{2}+\frac{1}{2}\zeta^{4}\eta^{2} \\
\end{pmatrix}$, \\
\end{longtable}
From the tensor product of the Siegel modular forms of weight 3/2 and 2, we find the Siegel modular forms of weight 7/2,
\begin{longtable}{l}
$Y^{(7/2)}_{\bm{1}_3} =
\begin{pmatrix}
Y^{(2)}_{\bm{1}_4}Y^{(3/2)}_{\bm{1}_7}
\end{pmatrix}_{\bm{1}_3} =
\begin{pmatrix}
-\sqrt{2}\zeta\eta\theta^{5}+\sqrt{2}\zeta\eta^{5}\theta-\frac{1}{\sqrt{2}}\zeta^{5}\theta^{2}+\frac{1}{\sqrt{2}}\zeta^{5}\eta^{2} \\
\end{pmatrix}$, \\
$Y^{(7/2)}_{\bm{2}_3} =
\begin{pmatrix}
Y^{(3/2)}_{\bm{\hat{3}}_5}Y^{(2)}_{\bm{3}'_6}
\end{pmatrix}_{\bm{2}_3} =
\begin{pmatrix}
\frac{1}{\sqrt{3}}\zeta\eta\theta^{5}-\frac{1}{\sqrt{3}}\zeta\eta^{5}\theta-\frac{1}{\sqrt{3}}\zeta^{5}\theta^{2}+\frac{1}{\sqrt{3}}\zeta^{5}\eta^{2} \\
-\frac{1}{4}\zeta\theta^{6}-\frac{5}{4}\zeta\eta^{2}\theta^{4}+\frac{5}{4}\zeta\eta^{4}\theta^{2}+\frac{1}{4}\zeta\eta^{6} \\
\end{pmatrix}$ \\
$Y^{(7/2)}_{\bm{3}_7} =
\begin{pmatrix}
Y^{(2)}_{\bm{6}_0}Y^{(3/2)}_{\bm{6}_3}
\end{pmatrix}_{\bm{3}_7} =
\begin{pmatrix}
\frac{1}{4}\zeta\theta^{6}-\frac{3}{4}\zeta\eta^{2}\theta^{4}+\frac{3}{4}\zeta\eta^{4}\theta^{2}-\frac{1}{4}\zeta\eta^{6} \\
\sqrt{2}\zeta^{3}\eta\theta^{3}-\sqrt{2}\zeta^{3}\eta^{3}\theta \\
-\frac{1}{\sqrt{2}}\zeta^{3}\theta^{4}+\frac{1}{\sqrt{2}}\zeta^{3}\eta^{4} \\
\end{pmatrix}$, \\
$Y^{(7/2)}_{\bm{3}'_5} =
\begin{pmatrix}
Y^{(3/2)}_{\bm{6}_3}Y^{(2)}_{\bm{3}_0}
\end{pmatrix}_{\bm{3}'_5} =
\begin{pmatrix}
\frac{1}{2}\zeta^{3}\theta^{4}-\zeta^{3}\eta^{2}\theta^{2}+\frac{1}{2}\zeta^{3}\eta^{4} \\
\frac{1}{2}\zeta^{2}\eta\theta^{4}-\zeta^{2}\eta^{3}\theta^{2}+\frac{1}{2}\zeta^{2}\eta^{5} \\
\frac{1}{2}\zeta^{2}\theta^{5}-\zeta^{2}\eta^{2}\theta^{3}+\frac{1}{2}\zeta^{2}\eta^{4}\theta \\
\end{pmatrix}$, \\
\end{longtable}
From the tensor product of the Siegel modular forms of weight 2, we find the Siegel modular forms of weight 4,
\begin{longtable}{l}
$Y^{(4)}_{\bm{1}_{0a}} =
\begin{pmatrix}
Y^{(2)}_{\bm{6}_0}Y^{(2)}_{\bm{6}_0}
\end{pmatrix}_{\bm{1}_0} =
\begin{pmatrix}
\frac{1}{4\sqrt{6}}\theta^{8}-\frac{1}{\sqrt{6}}\eta^{2}\theta^{6}+\frac{3}{2\sqrt{6}}\eta^{4}\theta^{4}-\frac{1}{\sqrt{6}}\eta^{6}\theta^{2}+\frac{1}{4\sqrt{6}}\eta^{8}-\frac{4\sqrt{6}}{3}\zeta^{4}\eta\theta^{3}-\frac{4\sqrt{6}}{3}\zeta^{4}\eta^{3}\theta \\
\end{pmatrix}$, \\
$Y^{(4)}_{\bm{1}_{0b}} =
\begin{pmatrix}
Y^{(2)}_{\bm{1}_4}Y^{(2)}_{\bm{1}_4}
\end{pmatrix}_{\bm{1}_0} =
\begin{pmatrix}
\frac{4}{3}\eta^{2}\theta^{6}+\frac{8}{3}\eta^{4}\theta^{4}+\frac{4}{3}\eta^{6}\theta^{2}+\frac{4}{3}\zeta^{4}\eta\theta^{3}+\frac{4}{3}\zeta^{4}\eta^{3}\theta+\frac{1}{3}\zeta^{8} \\
\end{pmatrix}$, \\
$Y^{(4)}_{\bm{2}_{0a}} =
\begin{pmatrix}
Y^{(2)}_{\bm{2}_0}Y^{(2)}_{\bm{1}_4}
\end{pmatrix}_{\bm{2}_0} =
\begin{pmatrix}
\frac{4}{3\sqrt{2}}\eta^{2}\theta^{6}+\frac{4\sqrt{2}}{3}\eta^{4}\theta^{4}+\frac{4}{3\sqrt{2}}\eta^{6}\theta^{2}-\frac{2}{3\sqrt{2}}\zeta^{4}\eta\theta^{3}-\frac{2}{3\sqrt{2}}\zeta^{4}\eta^{3}\theta-\frac{2}{3\sqrt{2}}\zeta^{8} \\
-\frac{1}{\sqrt{6}}\eta\theta^{7}-\frac{7}{\sqrt{6}}\eta^{3}\theta^{5}-\frac{7}{\sqrt{6}}\eta^{5}\theta^{3}-\frac{1}{\sqrt{6}}\eta^{7}\theta-\frac{1}{2\sqrt{6}}\zeta^{4}\theta^{4}-\frac{\sqrt{6}}{2}\zeta^{4}\eta^{2}\theta^{2}-\frac{1}{2\sqrt{6}}\zeta^{4}\eta^{4} \\
\end{pmatrix}$, \\
$Y^{(4)}_{\bm{2}_{0b}} =
\begin{pmatrix}
Y^{(2)}_{\bm{6}_0}Y^{(2)}_{\bm{6}_0}
\end{pmatrix}_{\bm{2}_0} =
\begin{pmatrix}
\frac{1}{4\sqrt{3}}\theta^{8}-\frac{1}{\sqrt{3}}\eta^{2}\theta^{6}+\frac{3}{2\sqrt{3}}\eta^{4}\theta^{4}-\frac{1}{\sqrt{3}}\eta^{6}\theta^{2}+\frac{1}{4\sqrt{3}}\eta^{8}+\frac{4\sqrt{3}}{3}\zeta^{4}\eta\theta^{3}+\frac{4\sqrt{3}}{3}\zeta^{4}\eta^{3}\theta \\
\zeta^{4}\theta^{4}+6\zeta^{4}\eta^{2}\theta^{2}+\zeta^{4}\eta^{4} \\
\end{pmatrix}$, \\
\end{longtable}
From the tensor product of the Siegel modular forms of weight 2 and 5/2, we find the Siegel modular forms of weight 9/2,
\begin{longtable}{l}
$Y^{(9/2)}_{\bm{1}_5} =
\begin{pmatrix}
Y^{(5/2)}_{\bm{\hat{3}}_7}Y^{(2)}_{\bm{3}'_6}
\end{pmatrix}_{\bm{1}_5} =
\begin{pmatrix}
\frac{3}{2\sqrt{6}}\zeta^{3}\theta^{6}-\frac{3\sqrt{6}}{4}\zeta^{3}\eta^{2}\theta^{4}+\frac{3\sqrt{6}}{4}\zeta^{3}\eta^{4}\theta^{2}-\frac{3}{2\sqrt{6}}\zeta^{3}\eta^{6} \\
\end{pmatrix}$, \\

$Y^{(9/2)}_{\bm{3}_{1a}} =
\begin{pmatrix}
Y^{(5/2)}_{\bm{3}'_{3b}}Y^{(2)}_{\bm{3}'_6}
\end{pmatrix}_{\bm{3}_1} =
\begin{pmatrix}
-\frac{1}{\sqrt{2}}\zeta^{7}\theta^{2}+\frac{1}{\sqrt{2}}\zeta^{7}\eta^{2} \\
\frac{3}{4}\zeta\eta\theta^{7}+\frac{7}{4}\zeta\eta^{3}\theta^{5}-\frac{7}{4}\zeta\eta^{5}\theta^{3}-\frac{3}{4}\zeta\eta^{7}\theta \\
\frac{1}{8}\zeta\theta^{8}+\frac{7}{4}\zeta\eta^{2}\theta^{6}-\frac{7}{4}\zeta\eta^{6}\theta^{2}-\frac{1}{8}\zeta\eta^{8} \\
\end{pmatrix}$, \\
$Y^{(9/2)}_{\bm{3}_{1b}} =
\begin{pmatrix}
Y^{(5/2)}_{\bm{3}'_{3a}}Y^{(2)}_{\bm{3}'_6}
\end{pmatrix}_{\bm{3}_1} =
\begin{pmatrix}
\sqrt{2}\zeta^{3}\eta\theta^{5}-\sqrt{2}\zeta^{3}\eta^{5}\theta \\
-\frac{1}{4}\zeta\eta\theta^{7}+\frac{3}{4}\zeta\eta^{3}\theta^{5}-\frac{3}{4}\zeta\eta^{5}\theta^{3}+\frac{1}{4}\zeta\eta^{7}\theta-\frac{1}{2}\zeta^{5}\theta^{4}+\frac{1}{2}\zeta^{5}\eta^{4} \\
\frac{1}{8}\zeta\theta^{8}-\frac{1}{4}\zeta\eta^{2}\theta^{6}+\frac{1}{4}\zeta\eta^{6}\theta^{2}-\frac{1}{8}\zeta\eta^{8}-\zeta^{5}\eta\theta^{3}+\zeta^{5}\eta^{3}\theta \\
\end{pmatrix}$, \\
\end{longtable}
From the tensor product of the Siegel modular forms of weight 5/2, we find the Siegel modular forms of weight 5,
\begin{longtable}{l}
$Y^{(5)}_{\bm{1}_2} =
\begin{pmatrix}
Y^{(5/2)}_{\bm{3}_5}Y^{(5/2)}_{\bm{3}_5}
\end{pmatrix}_{\bm{1}_2}$\\$ =
\begin{pmatrix}
\frac{1}{\sqrt{3}}\zeta^{2}\eta\theta^{7}-\frac{1}{\sqrt{3}}\zeta^{2}\eta^{3}\theta^{5}-\frac{1}{\sqrt{3}}\zeta^{2}\eta^{5}\theta^{3}+\frac{1}{\sqrt{3}}\zeta^{2}\eta^{7}\theta+\frac{1}{2\sqrt{3}}\zeta^{6}\theta^{4}-\frac{1}{\sqrt{3}}\zeta^{6}\eta^{2}\theta^{2}+\frac{1}{2\sqrt{3}}\zeta^{6}\eta^{4} \\
\end{pmatrix}$, \\

$Y^{(5)}_{\bm{2}_2} =
\begin{pmatrix}
Y^{(5/2)}_{\bm{\hat{3}}_7}Y^{(5/2)}_{\bm{3}'_7}
\end{pmatrix}_{\bm{2}_2}$\\$ =
\begin{pmatrix}
\frac{1}{\sqrt{2}}\zeta^{2}\eta\theta^{7}-\frac{1}{\sqrt{2}}\zeta^{2}\eta^{3}\theta^{5}-\frac{1}{\sqrt{2}}\zeta^{2}\eta^{5}\theta^{3}+\frac{1}{\sqrt{2}}\zeta^{2}\eta^{7}\theta-\frac{1}{\sqrt{2}}\zeta^{6}\theta^{4}+\sqrt{2}\zeta^{6}\eta^{2}\theta^{2}-\frac{1}{\sqrt{2}}\zeta^{6}\eta^{4} \\
-\frac{3}{4\sqrt{6}}\zeta^{2}\theta^{8}-\frac{\sqrt{6}}{2}\zeta^{2}\eta^{2}\theta^{6}+\frac{5\sqrt{6}}{4}\zeta^{2}\eta^{4}\theta^{4}-\frac{\sqrt{6}}{2}\zeta^{2}\eta^{6}\theta^{2}-\frac{3}{4\sqrt{6}}\zeta^{2}\eta^{8} \\
\end{pmatrix}$, \\

$Y^{(5)}_{\bm{3}_{2a}} =
\begin{pmatrix}
Y^{(5/2)}_{\bm{3}'_{3a}}Y^{(5/2)}_{\bm{3}'_7}
\end{pmatrix}_{\bm{3}_2}$\\$ =
\begin{pmatrix}
-\zeta^{2}\eta\theta^{7}-7\zeta^{2}\eta^{3}\theta^{5}-7\zeta^{2}\eta^{5}\theta^{3}-\zeta^{2}\eta^{7}\theta \\
-\frac{3}{8\sqrt{2}}\eta\theta^{9}-\frac{1}{2\sqrt{2}}\eta^{3}\theta^{7}+\frac{7\sqrt{2}}{8}\eta^{5}\theta^{5}-\frac{1}{2\sqrt{2}}\eta^{7}\theta^{3}-\frac{3}{8\sqrt{2}}\eta^{9}\theta-\sqrt{2}\zeta^{8}\eta\theta \\
\frac{1}{16\sqrt{2}}\theta^{10}+\frac{13}{16\sqrt{2}}\eta^{2}\theta^{8}-\frac{7}{8\sqrt{2}}\eta^{4}\theta^{6}-\frac{7}{8\sqrt{2}}\eta^{6}\theta^{4}+\frac{13}{16\sqrt{2}}\eta^{8}\theta^{2}+\frac{1}{16\sqrt{2}}\eta^{10}-\frac{1}{\sqrt{2}}\zeta^{8}\theta^{2}-\frac{1}{\sqrt{2}}\zeta^{8}\eta^{2} \\
\end{pmatrix}$, \\
$Y^{(5)}_{\bm{3}_{2b}} =
\begin{pmatrix}
Y^{(5/2)}_{\bm{6}_1}Y^{(5/2)}_{\bm{3}'_{3a}}
\end{pmatrix}_{\bm{3}_2}$\\
\end{longtable}


\section{Fermion mass matrices with $\widetilde{\Delta}(96)$ modular symmetry}
\label{app:ferm}

Here we summarize all possible fermion mass matrices in $\widetilde{\Delta}(96)$ modular symmetric flavor models.
Fermion mass matrices are defined by
\begin{align}
M^{jk} = \sum_{\bm{r}} \alpha_{\bm{r}_{q_Y}}^{(k_Y)} \left(Y_{\bm{r}_{q_Y}}^{(k_Y)} \otimes L^j \otimes R^k \right)_{\bm{1}_0}.
\end{align}
Left-handed fermion $L^j$ and right-handed fermion $R^k$ are assigned into three-dimensional representation of $\widetilde{\Delta}(96)$ and possess the weights $k_{Lj}$ and $k_{Rk}$, respectively.
Then structures of $M^{jk}$ are shown in Table \ref{tab:matrices}.
We have used the notation,
\begin{align}
q_{jk} \equiv -q_{Lj} - q_{Rk}, \quad k_{jk} \equiv - k_{Lj} - k_{Rk}.
\end{align}

\begin{longtable}{c}
\caption{Fermion mass matrices in the modular flavor model with $\widetilde{\Delta}(96)$ symmetry.}
\label{tab:matrices} \\\hline
$L^T = \begin{pmatrix}
Y_{\bm{1}_{q_{L1}}}^{(k_{L1})} &
Y_{\bm{1}_{q_{L2}}}^{(k_{L2})} &
Y_{\bm{1}_{q_{L3}}}^{(k_{L3})} \\
\end{pmatrix}, \quad
R^T = \begin{pmatrix}
Y_{\bm{1}_{q_{R1}}}^{(k_{R1})} &
Y_{\bm{1}_{q_{R2}}}^{(k_{R2})} &
Y_{\bm{1}_{q_{R3}}}^{(k_{R3})} \\
\end{pmatrix}$ \\ \hline
$M = \alpha_{\bm{1}_{q_{11}}}^{(k_{11})}\begin{pmatrix}
Y_{\bm{1}_{q_{11}}}^{(k_{11}),1} & 0 & 0 \\
0 & 0 & 0 \\
0 & 0 & 0 \\
\end{pmatrix}
+ \alpha_{\bm{1}_{q_{12}}}^{(k_{12})}\begin{pmatrix}
0 & Y_{\bm{1}_{q_{12}}}^{(k_{12}),1} & 0 \\
0 & 0 & 0 \\
0 & 0 & 0 \\
\end{pmatrix}
+ \alpha_{\bm{1}_{q_{13}}}^{(k_{13})}\begin{pmatrix}
0 & 0 & Y_{\bm{1}_{q_{13}}}^{(k_{13}),1} \\
0 & 0 & 0 \\
0 & 0 & 0 \\
\end{pmatrix} $\\
$+ \alpha_{\bm{1}_{q_{21}}}^{(k_{21})}\begin{pmatrix}
0 & 0 & 0 \\
Y_{\bm{1}_{q_{21}}}^{(k_{21}),1} & 0 & 0 \\
0 & 0 & 0 \\
\end{pmatrix}
+ \alpha_{\bm{1}_{q_{22}}}^{(k_{22})}\begin{pmatrix}
0 & 0 & 0 \\
0 & Y_{\bm{1}_{q_{22}}}^{(k_{22}),1} & 0 \\
0 & 0 & 0 \\
\end{pmatrix}
+ \alpha_{\bm{1}_{q_{23}}}^{(k_{23})}\begin{pmatrix}
0 & 0 & 0 \\
0 & 0 & Y_{\bm{1}_{q_{23}}}^{(k_{23}),1} \\
0 & 0 & 0 \\
\end{pmatrix} $\\
$+ \alpha_{\bm{1}_{q_{31}}}^{(k_{31})}\begin{pmatrix}
0 & 0 & 0 \\
0 & 0 & 0 \\
Y_{\bm{1}_{q_{31}}}^{(k_{31}),1} & 0 & 0 \\
\end{pmatrix}
+ \alpha_{\bm{1}_{q_{32}}}^{(k_{32})}\begin{pmatrix}
0 & 0 & 0 \\
0 & 0 & 0 \\
0 & Y_{\bm{1}_{q_{32}}}^{(k_{32}),1} & 0 \\
\end{pmatrix}
+ \alpha_{\bm{1}_{q_{33}}}^{(k_{33})}\begin{pmatrix}
0 & 0 & 0 \\
0 & 0 & 0 \\
0 & 0 & Y_{\bm{1}_{q_{33}}}^{(k_{33}),1} \\
\end{pmatrix} $\\
\hline
$L^T = \begin{pmatrix}
Y_{\bm{1}_{q_{L1}}}^{(k_{L1})} &
Y_{\bm{2}_{q_{L2}}}^{(k_{L2})} \\
\end{pmatrix}, \quad
R^T = \begin{pmatrix}
Y_{\bm{1}_{q_{R1}}}^{(k_{R1})} &
Y_{\bm{1}_{q_{R2}}}^{(k_{R2})} &
Y_{\bm{1}_{q_{R3}}}^{(k_{R3})} \\
\end{pmatrix}$ \\
\hline
\end{longtable}


\end{document}